\documentclass[prd,tightenlines,floatfix,showpacs,preprintnumbers,nofootinbib,eqsecnum,superscriptaddress,twocolumn]{revtex4-1}

\usepackage{color}      % Non-standard graphics support
\usepackage[dvips]{graphicx}    % Non-standard graphics support
\usepackage{amsbsy}     % Bolds (vectors, tensors, etc.)
\usepackage{amsfonts}       % AMS fonts
\usepackage{amsmath}        % AMS maths
\usepackage{amssymb}        % AMS symbols
\usepackage{color}
\usepackage{wasysym}
\usepackage{multirow}
\usepackage{mciteplus}
\usepackage{psfrag}
\usepackage{slashed}
\usepackage{cancel}

\usepackage{hyperref}

\newcommand{\MO}{{\tt micrOMEGAs}}

\newcommand{\FeynArts}{{\tt FeynArts}}
\newcommand{\FORM}{{\tt Form}}

\newcommand{\FeynCalc}{{\tt FeynCalc}}

\newcommand{\als}{\alpha_{\rm s}}
\newcommand{\as}{a_{\rm s}}
\newcommand{\nf}{n^{}_{\! f}}
\newcommand{\nfs}{n^{2}_{\! f}}
\newcommand{\nft}{n^{3}_{\! f}}
\newcommand{\NmLO}{N$^{\rm m}$LO}

\catcode`\@=11
\font\manfnt=manfnt
\def\Watchout{\@ifnextchar [{\W@tchout}{\W@tchout[1]}}
\def\W@tchout[#1]{{\manfnt\@tempcnta#1\relax%
  \@whilenum\@tempcnta>\z@\do{%
    \char"7F\hskip 0.3em\advance\@tempcnta\m@ne}}}
\let\foo\W@tchout
\def\dubious{\@ifnextchar[{\@dubious}{\@dubious[1]}}

\def\@dubious[#1]{%
  \color{red}\setbox\@tempboxa\hbox{\@W@tchout#1}
  \@tempdima\wd\@tempboxa
  \list{}{\leftmargin\@tempdima}\item[\hbox to 0pt{\hss\@W@tchout#1}]}
\def\@W@tchout#1{\W@tchout[#1]}
\catcode`\@=12

\begin{document}
%\begin{flushright}
\preprint{LAPTH-049/14, LCTS/2014-33, MS-TP-14-24}
%arXiv:yymm.nnnn [hep-ph]
%\end{flushright}
%\vspace{1cm}

\title{One-loop corrections to neutralino-stop coannihilation revisited}

%\received{XX XXXXX 2006}

\author{J.~Harz}
 \email{j.harz@ucl.ac.uk}
 \affiliation{
	Department of Physics and Astronomy, University College London, London WC1E 6BT, United Kingdom
  }

\author{B.~Herrmann}
 \email{herrmann@lapth.cnrs.fr}
 \affiliation{
	LAPTh, Universit\'e de Savoie, CNRS, 9 Chemin de Bellevue, B.P.\ 110, F-74941 Annecy-le-Vieux, France
  }

\author{M.~Klasen}
 \email{michael.klasen@uni-muenster.de}
 \affiliation{
	Institut f\"ur Theoretische Physik, Westf\"alische Wilhelms-Universit\"at M\"unster, Wilhelm-Klemm-Stra{\ss}e 9, D-48149 M\"unster, Germany
  }

\author{K.~Kova\v{r}\'ik}
 \email{karol.kovarik@uni-muenster.de}
 \affiliation{
	Institut f\"ur Theoretische Physik, Westf\"alische Wilhelms-Universit\"at M\"unster, Wilhelm-Klemm-Stra{\ss}e 9, D-48149 M\"unster, Germany
  }

\date{\today}

\begin{abstract}
We discuss the ${\cal O}(\alpha_s)$ supersymmetric QCD corrections to neutralino-stop coannihilation into a top quark and a gluon in the Minimal Supersymmetric Standard Model (MSSM). This particular channel can be numerically important in wide ranges of the MSSM parameter space with rather light stops. We discuss technical details such as the renormalization scheme and the phase-space slicing method with two cutoffs. We also comment on improvements with respect to earlier works on the given process. Further, we study for the first time the phenomenologically very interesting interplay of neutralino-stop coannihilation with neutralino-pair annihilation into quark pairs taking the full next-to-leading order SUSY-QCD corrections into account. We demonstrate that the numerical impact of these corrections on the total (co)annihilation cross section and finally on the theoretically predicted neutralino relic density is significant.
\end{abstract}


\maketitle

%%%%%%%%%%%%%%%%%%%%%%%%%%%%%%%%%%%%%%%%%%%%%%%%%%%%%%%%%%%%%%%%%%%%%%%%%%%%%%%
%%%%%%%%%%%%%%%%%%%%%%%%%%%%%%%%%%%%%%%%%%%%%%%%%%%%%%%%%%%%%%%%%%%%%%%%%%%%%%%

% =============================================================================
%!TEX root=paper.tex

\section{Introduction \label{Sec:Intro}}
% =============================================================================

While there is striking evidence for the existence of cold dark matter (CDM) in our Universe today, its exact nature remains one of the most important open questions of modern physics. Recent measurements of the cosmic microwave background (CMB) from the Planck satellite, combined with results from WMAP polarization data, have allowed to determine the dark matter relic density $\Omega_{\rm CDM}h^2$ of the Universe to precisely \cite{Planck, WMAP9}
\begin{equation}
	\Omega_{\mathrm{CDM}}h^2 = 0.1199 \pm 0.0027.
	\label{Eq:Planck}
\end{equation}

Beyond the Standard Model, a common assumption is that cold dark matter invokes a new particle, which is stable due to some $Z_2$ symmetry. The relic density can then be predicted based on cosmology and particle physics. Denoting the dark matter candidate by $\chi$, the theoretical value of its relic density ($\Omega_{\chi} h^2$) is given by
\begin{equation}
	\Omega_{\chi} h^2 ~=~ \frac{m_{\chi} n_{\chi}}{\rho_{\rm crit}} ,
\end{equation}
where $h$ is the Hubble constant, $m_{\chi}$ is the mass of the dark matter particle, and $\rho_{\rm crit}$ is the critical density of the Universe. Moreover, $n_{\chi}$ denotes today's number density of the dark matter particle in the Universe. This number can be obtained by solving the Boltzmann equation,
\begin{equation}
	\frac{\mathrm{d}n_\chi}{\mathrm{d}t} = -3 H n_\chi 
		- \left\langle\sigma_{\mathrm{ann}}v\right\rangle \Big[ n_\chi^2 
		- \left( n_\chi^{\mathrm{eq}} \right)^2 \Big],
	\label{Eq:Boltzmann}
\end{equation}
which describes the time evolution of the number density of a thermal relic in the Universe. The term proportional to the Hubble parameter $H$ describes the dilution due to the expansion of the Universe, while the second term on the right-hand side corresponds to (co)annihilation of the relic particle into Standard Model particles. It contains the (co)annihilation cross section \cite{Gondolo1991, Griest1991, Edsjo1997}
\begin{equation}
	\left\langle \sigma_{\mathrm{ann}}v\right \rangle = 
		\sum_{i,j} \langle \sigma_{ij}v_{ij} \rangle \frac{n_i^{\mathrm{eq}}}{n_\chi^{\mathrm{eq}}}
			\frac{n_j^{\mathrm{eq}}}{n_\chi^{\mathrm{eq}}},
	\label{Eq:Sigma}
\end{equation}
where the sum runs over all $Z_2$-odd particles of the theory. In this manner, Eq.\ (\ref{Eq:Sigma}) accounts for pair-annihilation of the dark matter particle as well as for coannihilation processes \cite{Salati1983} with other $Z_2$-odd particles. The equilibrium number densities appearing in Eqs.\ (\ref{Eq:Boltzmann}) and (\ref{Eq:Sigma}) are related to the mass of the corresponding particle as well as to the temperature $T$ through $n_i^{\mathrm{eq}} \sim \exp \left\{ -m_i / T \right\}$.
As a consequence, the ratios of the number densities appearing in Eq.\ (\ref{Eq:Sigma}) are given by
\begin{equation}
	\frac{n_i^{\mathrm{eq}}}{n_\chi^{\mathrm{eq}}} ~\sim~ \exp\left\{ -\frac{m_i-m_\chi}{T} \right\}.
	\label{Eq:Suppression}
\end{equation}
This shows that coannihilation processes are relevant in scenarios where another particle is almost degenerate in mass with the dark matter particle. 

In the present paper, we focus on the case of the Minimal Supersymmetric Standard Model (MSSM), in which the lightest neutralino is the most popular candidate for cold dark matter. The neutralino is a mixture of the bino $\tilde{B}$, the wino $\tilde{W}$, and the higgsinos $\tilde{H}_1$ and $\tilde{H}_2$,
\begin{equation}
	\tilde{\chi}_1^0 ~=~ Z_{1\tilde{B}} \tilde{B} + Z_{1\tilde{W}} \tilde{W} + Z_{1\tilde{H}_1} \tilde{H}_1 + Z_{1\tilde{H}_2} \tilde{H}_2 \, .
\end{equation}
In case of a light stop, its coannihilation processes can be numerically dominant in the calculation of the cross section \cite{Boehm2000, Ellis2001}.

The parameters of the Boltzmann equation are affected by theoretical uncertainties, which have to be reduced in order to meet the experimental precision of Eq.\ (\ref{Eq:Planck}). On the cosmology side, e.g., variations in the Hubble expansion rate or altered assumptions on the underlying cosmological model give rise to uncertainties in the relic density prediction \cite{Hamann2007,Arbey2008, Arbey2009}. On the particle physics side, the main uncertainty resides in the calculation of the (co)annihilation cross sections $\sigma_{ij}$ appearing in Eq.\ (\ref{Eq:Sigma}). In publicly available tools such as {\tt micrOMEGAs} \cite{micrOMEGAs} or {\tt DarkSUSY} \cite{DarkSUSY}, these cross sections are evaluated only at the tree level taking into account effective quark masses and running couplings for certain cases. However, it is well known, that higher-order corrections can have a sizable impact on such cross sections and thus on the theory prediction of the relic density. 

This has been explicitly shown for different annihilation channels and scenarios with the common conclusion that the impact of the higher-order corrections on the relic density can be numerically more important than the current experimental uncertainty of Eq.\ (\ref{Eq:Planck}). Previous studies include neutralino pair-annihilation into quark-antiquark pairs \cite{DMNLO_AFunnel, DMNLO_mSUGRA, DMNLO_NUHM, DMNLO_ChiChi} and electroweak final states \cite{Sloops2007, Sloops2009, Sloops2010, Sloops2011, EffCouplings, SloopsEff2014} as well as coannihilations of the lighter neutralinos and charginos \cite{DMNLO_ChiChi}. Neutralino-stop coannihilation with Higgs or electroweak vector bosons in the final state has also been studied at the one-loop level \cite{DMNLO_Stop1}.

It is the aim of the present paper to extend and to combine the already existing analyses. With respect to our previous paper 
\cite{DMNLO_Stop1}, we now also include the neutralino-stop coannihilation with a gluon and a top-quark in the final state.
This final state was already considered in the analysis of Ref.\ \cite{Freitas2007}, although no further details were given concerning the corresponding calculation. Moreover, Ref.\ \cite{Freitas2007} focused on a rather special case of a bino-like neutralino, which coannihilates exclusively with a right-handed stop according to $\tilde{B} \tilde{t}_R \to t g$ and $\tilde{B} \tilde{t}_R \to b W^+$.

Our analysis extends Ref.\ \cite{Freitas2007} in several important aspects. First, we discuss in detail the treatment of the arising ultraviolet (UV) and infrared (IR) divergences in the process with a gluon in the final state, e.g., non-trivial issues like the renormalization of the strong coupling constant $\alpha_s$ as well as the phase-space slicing method, which is applied in order to cancel the infrared divergences and to properly evaluate the real emission cross sections. Moreover, our analysis is general and remains valid when the neutralino has sizable admixtures of wino and higgsinos altering its couplings and annihilation channels. Also, the lightest stop is likely to be a mixture of the left- and right-handed superpartners of the top quark, and a large mixing in the stop sector \cite{Haber1996, Badziak2012} is often required for a Higgs mass of about 125 GeV \cite{ATLAS2014, CMS2014}. Moreover, all corresponding final states with electroweak vector bosons have been taken into account.

Furthermore, we study for the first time the phenomenologically very realistic and interesting interplay of neutralino pair annihilation into quark pairs \cite{DMNLO_ChiChi} and neutralino-stop coannihilation processes. Apart from the newly added stop-neutralino coannihilation into a gluon and a top quark, we include the coannihilation into a top quark and a Higgs boson which becomes significant if one attempts to achieve a Higgs mass of around 125 GeV by a large trilinear coupling in the stop sector as shown in Ref.\ \cite{DMNLO_Stop1}.

Our paper is organized as follows: In Sec.\ \ref{Sec:Technical} we present our calculation and discuss technical details such as the renormalization scheme and the infrared treatment. Numerical results for the annihilation cross section and the neutralino relic density are shown in Sec.\ \ref{Sec:Results}, which includes also a phenomenological discussion of the results. Finally, conclusions are given in Sec.\ \ref{Sec:Conclusion}.

% =============================================================================
%!TEX root=paper.tex

\section{Technical details \label{Sec:Technical}}
% =============================================================================

The analysis presented in this paper involves a calculation of the coannihilation cross section at next-to-leading order in the strong coupling constant. In order to provide predictions for the relic density up to the next-to-leading order, one has to consistently calculate all relevant processes up to that same order. In our case, we consider the pair-annihilation of neutralinos into heavy quarks and coannihilation of the lightest neutralino and scalar top quarks into a quark and an electroweak gauge boson, a Higgs boson, or a gluon.

Most of these processes have been separately analyzed in our previous work \cite{DMNLO_ChiChi, DMNLO_Stop1}. Here, we combine them for the first time within a single analysis and further add a new and important coannihilation process, namely the process with a gluon in the final state. Next-to-leading order corrections to these processes involve one-loop diagrams, which are ultraviolet and infrared divergent. The UV divergences are cancelled by renormalization, while the IR divergences vanish when including $2\rightarrow 3$ processes with an additional parton in the final state.

\begin{figure}
 	\includegraphics[scale=0.7]{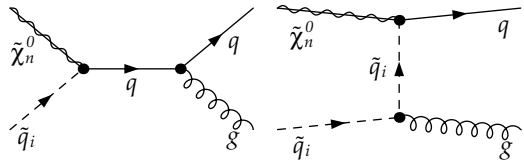}
 	\caption{Leading-order Feynman diagrams for neutralino-squark coannihilation into a quark and a gluon.}
 	\label{Fig:TreeDiagrams}
\end{figure}

\begin{figure*}
	\includegraphics[scale=1.1]{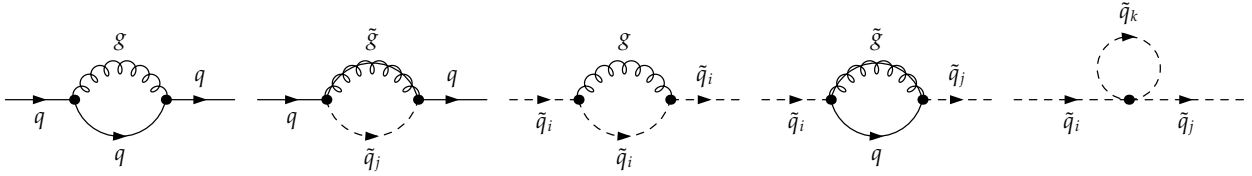}
	\caption{Quark and squark self-energies at the one-loop level contributing through the propagators in the $s$- and $t$-channel, respectively.}
	\label{Fig:SelfEnergies_squark}
\end{figure*}

\begin{figure*}
	\includegraphics[scale=1.]{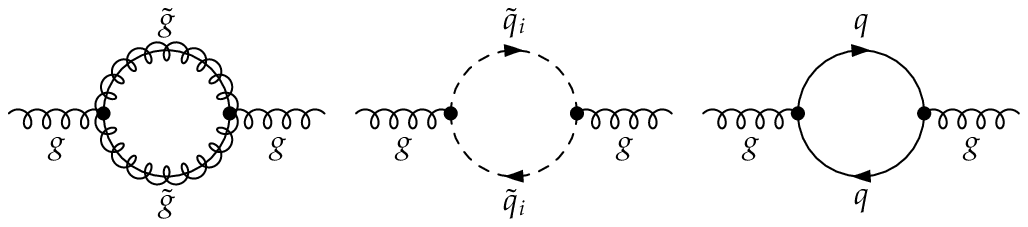}
	\includegraphics[scale=1.]{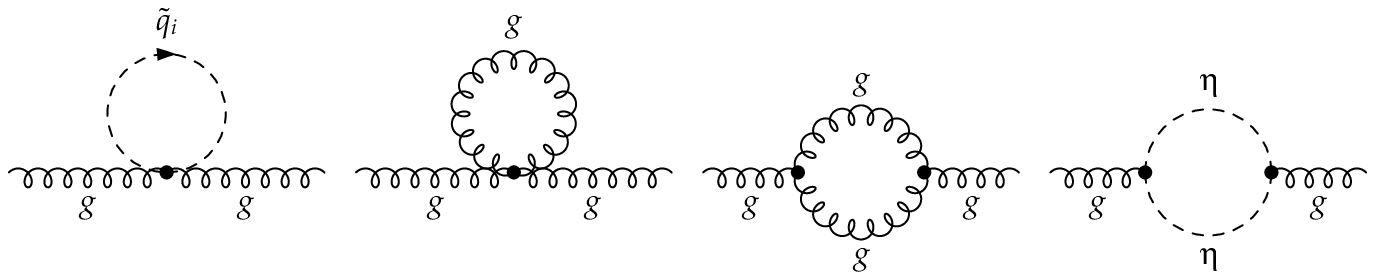}
	\caption{Gluon self-energies at the one-loop level leading to the gluon wave-function renormalization constant $\delta Z_g$, which enters the counterterm of the strong coupling constant as well as directly the counterterms to the quark-quark-gluon and squark-squark-gluon coupling.}
	\label{Fig:SelfEnergies}
\end{figure*}

\begin{figure*}
	\includegraphics[scale=0.92]{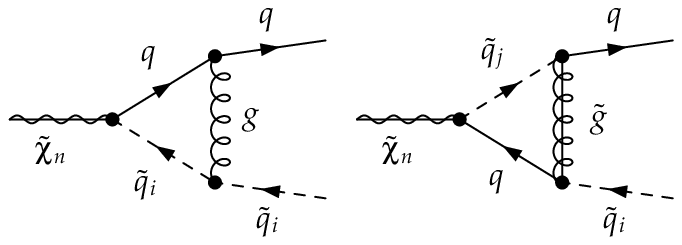}\\
	\includegraphics[scale=0.92]{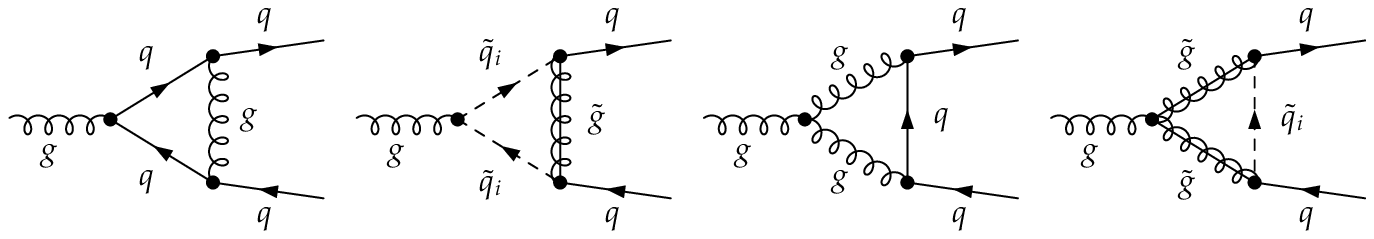}\\
	\includegraphics[scale=0.92]{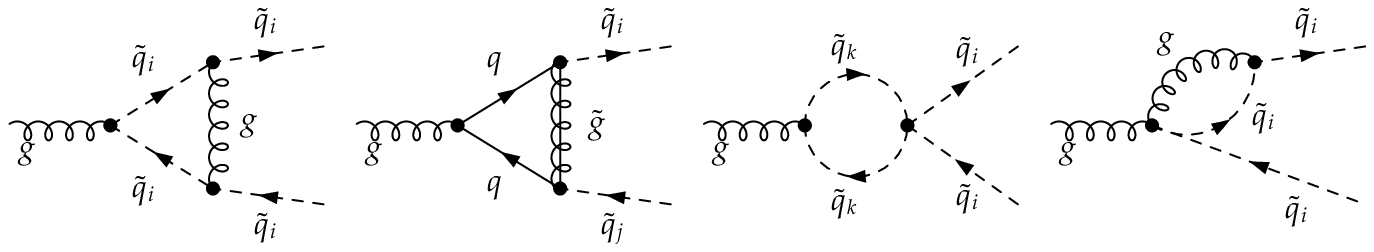}\\
	\includegraphics[scale=0.92]{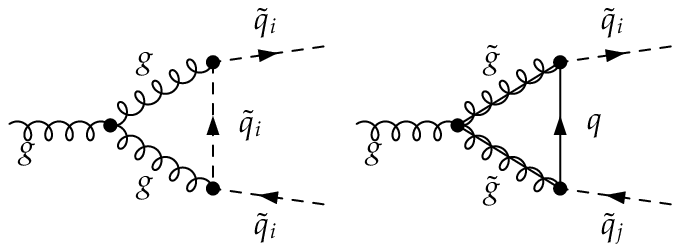}
	\caption{Vertex corrections at the one-loop level contributing to neutralino-squark coannihilation into a gluon in the final state.
	The first row arises from the neutralino-squark-quark-coupling, the second row from the gluon-quark-quark-coupling, and the third 
	and fourth rows from the gluon-squark-squark-coupling. The first diagram of the second and third row, respectively, give rise to an 
infrared single pole (soft). The diagrams with a gluon-gluon-gluon-vertex, however, lead to an infrared double pole (soft-collinear).}
	\label{Fig:VertexCorrections}
\end{figure*}

\begin{figure*}
      \begin{minipage}{1.0\linewidth}
	\includegraphics[scale=1.15]{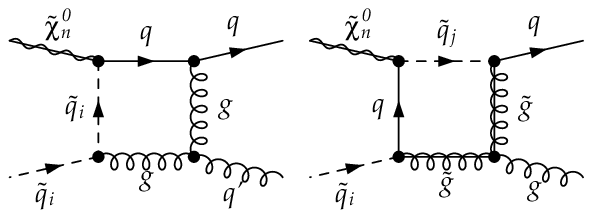}\\
	\includegraphics[scale=1.15]{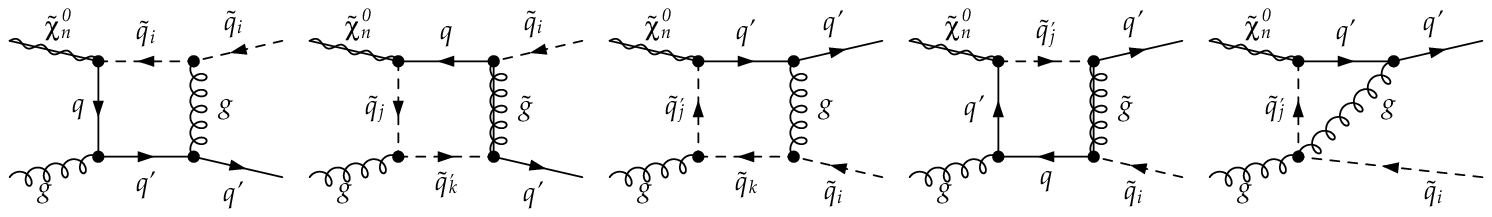}
	\end{minipage}
	\caption{Box diagrams at the one-loop level contributing to neutralino-squark coannihilation into a gluon. The first diagram in the first row leads to an infrared double pole (soft-collinear), the first and third diagrams in the second row give rise to an infrared single pole (soft).}
	\label{Fig:BoxDiagrams}
\end{figure*}

\begin{figure*}
\begin{minipage}{1.0\linewidth}
        \includegraphics[scale=0.92]{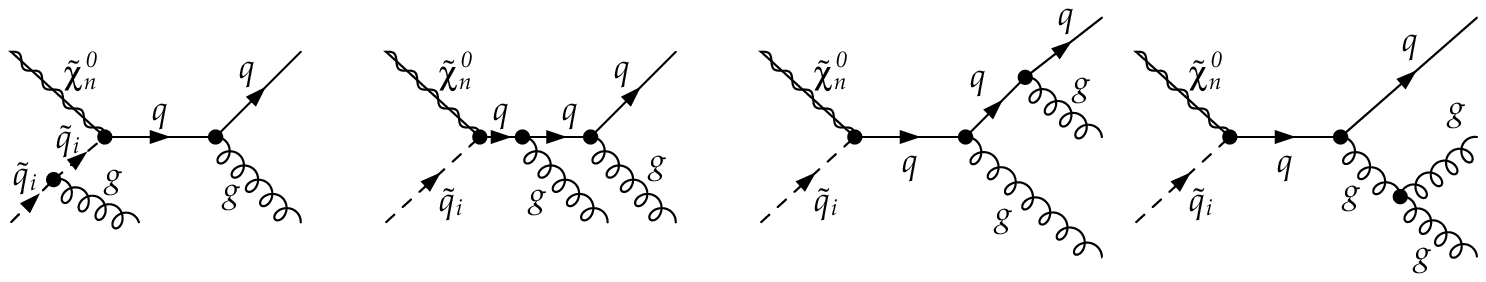}\\
	\includegraphics[scale=0.92]{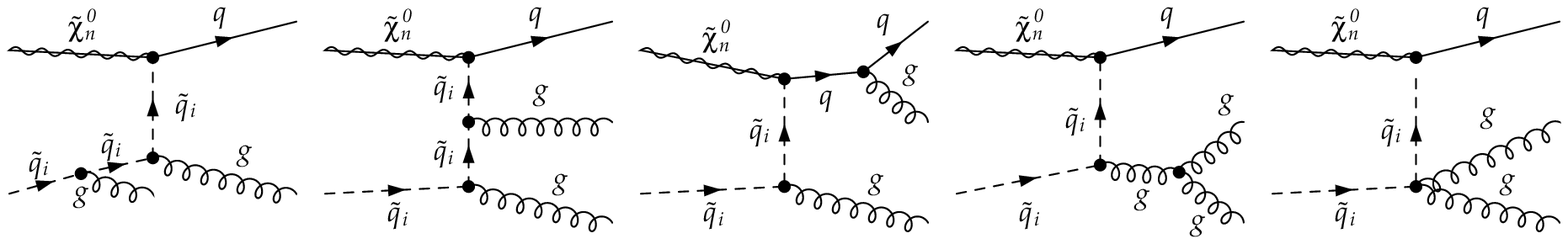}\\
	\includegraphics[scale=0.92]{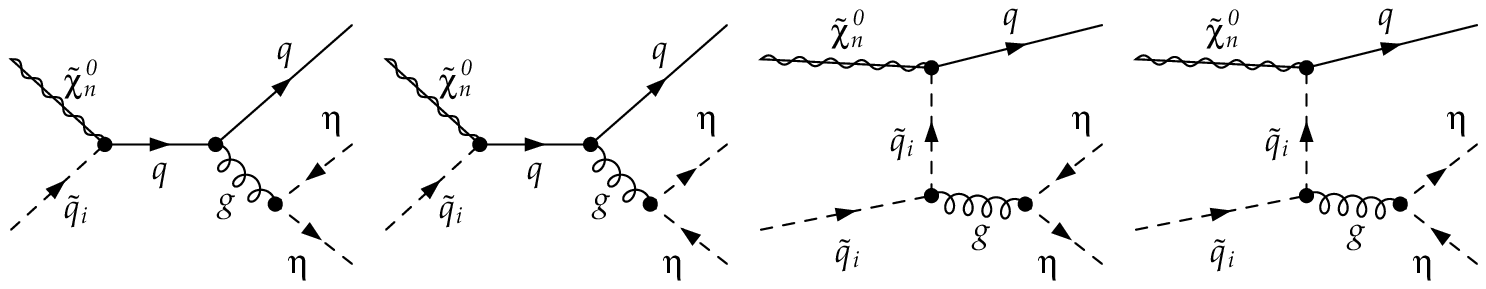}\\
	\includegraphics[scale=0.92]{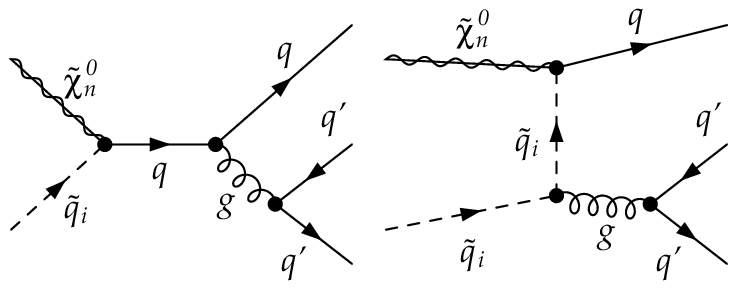}
	\end{minipage}
	\caption{Real gluon emission diagrams contributing at the next-to-leading order to neutralino-squark coannihilation with a gluon in the final state. The first and second rows show the usual emission from the initial state stop, the final state top and gluon as well as from the corresponding propagator particle (top in case of $s$-channel, stop for the $t$-channel). For the $t$-channel diagram also radiation off the squark-squark-gluon vertex is possible. In the third row the diagrams containing ghosts are depicted, which are needed in order to maintain gauge invariance. Further, light quark emission off the final state gluon is possible (fourth row).}
	\label{Fig:RealEmission}
\end{figure*}

In the following, we will discuss additions which have to be made to the renormalization scheme and to the treatment of IR divergences in order to treat the coannihilation processes with a gluon in the final state (see Fig.\ \ref{Fig:TreeDiagrams}). The necessary one-loop diagrams and $2 \rightarrow 3$ processes for a gluon in the final state are depicted in Figs.\ \ref{Fig:SelfEnergies_squark} -- \ref{Fig:RealEmission}. All diagrams have been calculated by using the publicly available tools {\FeynArts} \cite{FeynArts}, {\FeynCalc} \cite{FeynCalc} and {\FORM} \cite{FORM}. In order to regularize the occurring UV divergences, we calculate in $D = 4 - 2 \varepsilon$ dimensions and use the dimensional reduction ($\overline{\tt DR}$) scheme, which preserves supersymmetry in contrast to the conventional minimal subtraction scheme ($\overline{\tt MS}$). The whole calculation is performed in the 't Hooft-Feynman gauge, which means we have included also external Faddeev-Popov ghosts.

In comparison to our previous calculation in Ref.~\cite{DMNLO_Stop1}, the coannihilation
process $\tilde{\chi}^0_1\, \tilde{t}_1\rightarrow tg$ involves diagrams with a more complicated gauge structure. As a result
several different color factors appear. Therefore, in every contribution, virtual or real, we identify the gauge invariant color 
classes and treat each color class independently. 

The color class proportional to the Casimir operator eigenvalue $C_F$ is similar to our previous calculation, where instead of the 
gluon, we had a color singlet electroweak boson in the final state. All ultraviolet and infrared divergences in that case can be 
treated analogously as in Ref.~\cite{DMNLO_Stop1}. 

The additional complications come from the contributions proportional to the other Casimir operator eigenvalue $C_A$. This class 
of diagrams includes diagrams with a triple-gluon coupling which introduce collinear infrared divergences in addition to the infrared 
soft and ultraviolet divergences. This causes a much more intricate divergence structure and requires a dedicated treatment, especially 
of the infrared divergences.  

There is yet another small invariant class of diagrams which is proportional to the $SU(3)$ invariant $T_f$ and is connected
to a closed fermion loop. If the fermion happens to be a light quark, these contributions are also infrared divergent.

% -----------------------------------------------------------------------------
\subsection{Renormalization}
% -----------------------------------------------------------------------------
%
Ultraviolet divergences that arise when calculating loop diagrams can be removed by a suitable redefinition of parameters and fields. In Ref.\ \cite{DMNLO_Stop1}, we have proposed a renormalization scheme suitable for all relevant annihilation and coannihilation processes. To circumvent known issues of previous on-shell and $\overline{\tt DR}$ renormalization schemes, we have put forward a mixed scheme where some input parameters are defined on-shell ($m_t$, $m_{\tilde{t}_1}$, $m_{\tilde{b}_1}$, $m_{\tilde{b}_2}$) and others are defined in the $\overline{\tt DR}$ renormalization scheme ($m_b$, $A_t$, $A_b$). The renormalization and factorization scales are set to 1~TeV, which corresponds to the scale at which the $\overline{\tt DR}$-input values are given \cite{SPA}. In our last paper \cite{DMNLO_ChiChi}, we have improved on the treatment of heavy quark masses and their Yukawa couplings.

Here, we will focus on the renormalization of the strong coupling constant $\alpha_s$ and of the gluon field, which are necessary ingredients to render finite the one-loop cross section of the newly added coannihilation process with a gluon in the final state.

% -----------------------------------------------------------------------------
\subsubsection{Gluon wave-function renormalization}
% -----------------------------------------------------------------------------

Having an external gluon requires the renormalization of the gluon field. We renormalize the gluon field by rescaling it using a wave-function renormalization constant $Z_g$,
\begin{equation}
	A^\mu\rightarrow \sqrt{Z_g}A^\mu = \left[ 1+\frac12 \delta Z_g \right] A^\mu .
\end{equation}
The wave-function renormalization constant $Z_g$ is connected to the general two-point Green's function of a vector field which can be parametrized as
\begin{equation}
	\Pi^{\mu\nu}(k^2) = \left[ g^{\mu\nu}-\frac{k^\mu k^\nu}{k^2} \right] \Pi^T(k^2) + \frac{k^\mu k^\nu}{k^2}\Pi^L(k^2),
\end{equation}
where $\Pi^T$ and $\Pi^L$ are the transverse and longitudinal form factors which receive contributions from all diagrams in Fig.\ \ref{Fig:SelfEnergies}. By requiring that the gluon propagator has a unit residue even at the one-loop level, we obtain the following expression for the renormalization constant:
\begin{equation}
	\delta Z_g = -\Re \left\{ \left.\frac{\partial \Pi^T (k^2)}{\partial k^2}\right|_{k^2=0}\right\} .
\end{equation}
This renormalization constant is both ultraviolet and infrared divergent. The ultraviolet divergence of the constant can be written as
\begin{equation}
	\delta Z_g^{\tt UV} = \frac{\alpha_s}{4\pi}\frac{1}{\varepsilon} \Big[ C_A - 2 \, T_f \, n_q \Big]\,,
\end{equation}
where the ultraviolet divergence manifests itself as a pole $\frac{1}{\varepsilon}$, $n_q$ indicates the number of all quarks,
and the constants $C_A$ and $T_f$ are the conventional $SU(3)$ invariants 
\begin{equation}
	C_A = N_C = 3\,,\qquad T_f = \frac12\,.
\end{equation}
The infrared divergence of the gluon wave-function renormalization constant is caused by an exchange of a massless gluon in the one-loop diagrams contributing to the Green's function and also by a collinear radiation of a massless particle (gluon, Faddeev-Popov ghost, or a massless quark) from another massless particle. The infrared divergent part of the gluon wave-function renormalization constant is 
\begin{equation}
	\delta Z_g^{\tt IR} = \frac{\alpha_s}{4\pi}\frac{1}{\varepsilon}\biggr[ -\frac{5}{3}C_A+\frac{4}{3}\, T_f\, n_{q^{\prime}} \biggr],
\end{equation}
where $n_{q^{\prime}}$ is the number of effectively massless quarks.

% -----------------------------------------------------------------------------
\subsubsection{Renormalization of $\alpha_s$}
% -----------------------------------------------------------------------------

In order to specify the renormalization of any parameter, one needs to give its definition which is unavoidably tied to its input value and the corresponding counter\-term resulting from the definition. Throughout our calculation, we use the strong coupling constant defined in the $\overline{\tt DR}$-scheme in the MSSM. Its value in this scheme is scale dependent and has to be obtained from the conventional value extracted from experimental data which is defined in the $\overline{\tt MS}$-scheme in the Standard Model and extracted at the scale $Q=m_Z$ of the mass of the $Z$-boson. The procedure to obtain the desired value of $\alpha_s$ can be sketched as follows:
\begin{widetext}
\begin{equation}
	\alpha_s^{\tt \overline{MS},SM,n_f=5}(m_Z^2)\overset{(1)}{\longrightarrow} \alpha_s^{\tt 
	\overline{MS},SM,n_f=5}(Q^2)\overset{(2)}{\longrightarrow} \alpha_s^{\tt \overline{DR},SM,n_f=5}(Q^2)
	\overset{(3)}{\longrightarrow} \alpha_s^{\tt \overline{DR},MSSM,n_f=6}(Q^2)\,.
	\label{eq:trans_as}
\end{equation}
\end{widetext}
There are several steps in the transformation which can be performed in different order: We have to change the scale of the coupling constant, transform the coupling from the $\overline{\tt MS}$-scheme to the $\overline{\tt DR}$-scheme and add effects of heavy particles such as the top quark and all colored supersymmetric particles (for a discussion of different approaches see \cite{Harlander:2007wh}). We chose the following sequence of transformations:

In the first step, we use the well-known scale dependence of the strong coupling constant in the Standard Model in the $\overline{\tt MS}$-scheme which at \NmLO\ is given by
\begin{equation}
  \frac{{\rm d}\, \as}{{\rm d} \log \mu_r^2} \: = \: \beta_{\:\!\rm N^mLO}(\as)
  \: = \: - \sum_{k=0}^m \, \as^{k+2} \,\beta_k \: ,
\end{equation}
with the shorthand notation $\as \equiv \als^{\tt\overline{MS},SM,n_f=5}/4\pi$. The renormalization scale is denoted by $\mu_r$ and $\nf$ stands for the number of effectively massless quark flavors, which is set to $n_f=5$ in our case. 

We use the expansion coefficients $\beta_k$ of the $\beta$-function of QCD at $k=3$, i.e., at N$^3$LO \cite{Vermaseren:1997fq} 
\begin{eqnarray}
  \beta_0 &\, =\, & \: 11\: - \: 2/3\, \nf\,,
  \nonumber \\
  \beta_1 &\, =\, & 102 - 38/3\, \nf\,,
  \\
  \beta_2 &\, =\, & \, 2857/2\, - 5033/18\, \nf + \,325/54\, \nfs\,,
  \nonumber \\
  \beta_3 &\, =\, & 29243.0 - \: 6946.30\: \nf + 405.089\, \nfs
                  + 1093/729\, \nft  .\nonumber
\end{eqnarray}
After we have shifted the scale using three-loop renormalization group equations from $Q^2=m_Z^2$ to the final scale 
$Q_{\tt fin}^2 = 1~\mathrm{TeV}^2$, all remaining steps of Eq.\ (\ref{eq:trans_as}) are performed at this scale.

The next step is rather more involved, as it requires a consistent definition of $\alpha_s$ in the dimensional reduction scheme within QCD without supersymmetry. The two-loop relation between the two definitions of the strong coupling constant can be written as \cite{Harlander:2005wm}
\begin{align}
  \alpha_s^{\tt \overline{DR}} = \alpha_s^{\tt \overline{MS}} &\left[ 1 + \frac{\alpha_s^{\tt \overline{MS}}}{\pi} \frac{C_A}{12} \right. \\
  &+ \left. \left( \frac{\alpha_s^{\tt \overline{MS}}}{\pi}\right)^{\!\! 2} \left(
  \frac{11}{72} C_A^2 - \frac{1}{8} C_F T n_f \right) \right]\!,\quad
  \nonumber
  \label{eq::asMS2DR}
\end{align}
where it is understood that both coupling constants are evaluated at the same scale, in the Standard Model and with $n_f=5$. 

The last step is to obtain a coupling constant in the MSSM, taking into account the existence of heavier supersymmetric particles which alter the scale dependence. The effects of the heavy top quark are also taken into account in this step in parallel with all other particles so that we obtain $\alpha_s^{\tt \overline{DR},MSSM,n_f=6}\equiv \alpha_s^{\tt full}$. The relation between the final strong coupling and the Standard Model one can be cast into the form
\begin{equation}
	\alpha_s^{\tt full} = \alpha_s^{\tt \overline{DR}} \left[ 1 - \frac{\alpha_s^{\tt \overline{DR}}}{\pi}\zeta_{1} + 
	\left(\frac{\alpha_s^{\tt \overline{DR}}}{\pi}\right)^{\!\! 2} \big(2 \zeta_{1}^2 - \zeta_{2}\big)\right] \,,
\end{equation}
where the first-order decoupling coefficient $\zeta_1$ is given by
\begin{equation} 
  \zeta_1 = 
  - \frac{1}{6} \log\frac{Q_{\tt fin}^2}{m_t^2} -  \frac{1}{24} \sum_{q}\sum_{i=1,2} \log\frac{Q_{\tt fin}^2}{m_{\tilde{q}_i}^2} - \frac{1}{2}\log\frac{Q_{\tt fin}^2}{m_{\tilde{g}}^2} \,.
  \label{eq::zetag1l}
\end{equation}
The explicit result for the second-order decoupling coefficient is too long to be shown here in its entirety. For all the details and the results for some special cases, we refer the reader to Ref.\ \cite{Bauer:2008bj}.

After we have established the value of the strong coupling constant, the corresponding counterterm remains to be specified. The counterterm of $\alpha_s$ in the MSSM in the $\overline{\tt DR}$-scheme is 
\begin{equation}
	\delta\alpha_s = \frac{\alpha_s}{8\pi}\Delta \biggr[ n_q - 3\,C_A \biggr] ,
\end{equation}
where $\Delta = \frac{1}{\varepsilon} -\gamma_E + \log 4\pi$.

With the treatment described above, a UV-finite calculation is achieved, which has been validated by various consistency checks.

% -----------------------------------------------------------------------------
\subsection{Phase-space slicing}
% -----------------------------------------------------------------------------
%
The infrared (IR) divergences occurring in the virtual corrections are cancelled by including the real emission processes
\begin{equation}
	\tilde{\chi}^0_1(p_1) + \tilde{t}_1(p_2)\rightarrow t(p_3)+g(p_4)+g(k)
\end{equation}
and
\begin{equation}
	\tilde{\chi}^0_1(p_1) + \tilde{t}_1(p_2)\rightarrow t(p_3)+q(p_4)+\bar{q}(k).
\end{equation}
The corresponding diagrams are depicted in Fig.\ \ref{Fig:RealEmission}. Whereas the infrared divergences of the virtual part can  
be explicitly isolated in $D=4-2\varepsilon$ dimensions, those in the real part result from the integration over the gluon phase space. 

\begin{figure}
	\includegraphics[scale=0.4]{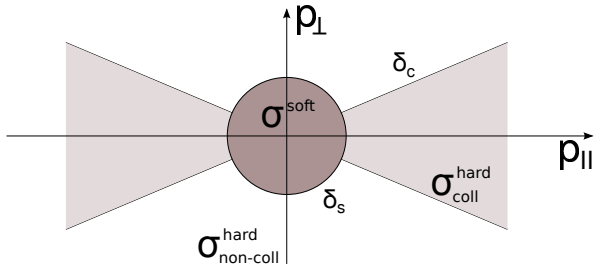}
	\caption{Schematic picture of the sliced phase space with the two cutoffs $\delta_s$ and $\delta_c$.}
	\label{Fig:RealPhaseSpace}
\end{figure}

In contrast to neutralino-stop coannihilation with a Higgs or electroweak vector boson in the final state, where only soft divergences  appear (see Ref.\ \cite{DMNLO_Stop1}), in the case of a gluon in the final state additional collinear divergences have to be considered. Therefore, a simple phase-space slicing method with just one cutoff on the gluon energy is not sufficient any more, and the method has to be extended to two cutoffs in order to distinguish between the soft (S), hard collinear (HC), and hard non-collinear (H$\overline{\mathrm{C}}$) regions of parameter space (see Fig.\ \ref{Fig:RealPhaseSpace}). To do so, we use the two-cutoff phase-space slicing method as introduced and discussed in Ref.\ \cite{HarrisOwens}. The first cutoff $\delta_s$ is applied on the gluon energy to distinguish between the soft and hard phase space. The second cutoff $\delta_c$ is used to separate the hard collinear and hard non-collinear phase space. In this way, the full $2 \rightarrow 3$ cross section is split into three parts,
\begin{equation}
	\sigma^{2\rightarrow 3}_{\rm full} = \sigma_{\rm S} + \sigma_{\rm HC} + \sigma_{{\rm H}\overline{\mathrm{C}}} .
\end{equation}

Having used the cutoffs $\delta_s$ and $\delta_c$ to decompose the $2 \rightarrow 3$ phase space into the three different regions as shown in Fig.\ \ref{Fig:RealPhaseSpace}, we can integrate now the full $2 \rightarrow 3$ matrix element in the hard non-collinear (and thus finite) region numerically. 
At this point, one should note that we apply the cutoffs only selectively on those subclasses of diagrams which would give rise to divergences when integrating over the squared amplitudes.
The whole list of squared amplitudes and the corresponding cuts are given in Tab.\ \ref{Tab:DivergentCases}.

In contrast to the hard non-collinear region, in the soft or collinear parts of the phase space, where the matrix element is divergent, we use the eikonal or leading collinear pole approximation of the full matrix element. Both approximations rely on the factorization of the differential $2 \to 3$ cross section in terms of the $2 \to 2$ matrix element according to
\begin{equation}
	\left( \frac{d\sigma}{d\Omega} \right)_{\textnormal{S,HC}} = 
	\left( \frac{d\sigma}{d\Omega} \right)_{\textnormal{tree}} \times F_{\textnormal{S,HC}},
\end{equation}
where the factors $F_{\textnormal{S,HC}}$ contain all infrared poles isolated with the help of the applied approximations. Tab.\ \ref{Tab:DivergentCases} gives an overview over the different divergent cases and categorizes the squared amplitudes arising from the different diagrams of Fig.\ \ref{Fig:RealEmission}. 

\begin{table*}
	\begin{tabular}{|l|c|l|}
	\hline
		& ~ Condition for pure hard $2\rightarrow 3$ processes ~ & Squared amplitudes \\
		\hline
Convergent & & SiTv, SpTv, SfTv, SgTv \\
 &    & TiTv, TpTv, TfTv, TgTv, TvTv \\
 \hline
Soft & $x_2 > 2 \delta_s / \sqrt{s}$ ~~and~~ $x_3 > 2 \delta_s / \sqrt{s}$ & SiSi, SiSp, SiSf, SpSp, SpSf, SfSf\\
 & & SiTi, SiTp, SpTi, SiTf, TfTi, SpTp, SpTf, SfTp, SfTf\\
 & & TiTi, TiTp, TiTf, TpTp, TpTf, TfTf\\
 \hline
Soft-collinear & $x_1 < 1 + \mu_3^2 - \delta_c/s$ ~~and~~ $x_2 > 2 \delta_s / \sqrt{s}$ & SiSg, SpSg, SfSg\\
 & ~~and~~ $x_3 > 2 \delta_s / \sqrt{s}$& SiTg, SpTg, SfTg, TiSg, TpSg, TfSg\\
  & & TiTg, TpTg, TfTg\\
   \hline
Hard-collinear & $x_1 < 1 + \mu_3^2 - \delta_c/s$ & SgSg, S$\eta$S$\eta$, SqSq\\
& & SgTg, S$\eta$T$\eta$, SqTq\\
& & TgTg, T$\eta$T$\eta$, TqTq\\
		\hline
	\end{tabular}
	\caption{Overview over all squared matrix elements corresponding to diagrams in Fig.\ \ref{Fig:RealEmission} together with the classification of their divergent behavior (convergent, soft, soft-collinear, hard-collinear) and the corresponding cuts imposed on the integration to avoid the divergences. In the last column the corresponding squared amplitudes are listed indicating if the diagrams squared correspond to the $S$-channel or the $T$-channel and indicating also the particle which radiates the gluon in each of the diagrams: i = initial state squark, p = propagator, f = final state quark, g = gluon, v = vertex; or in case of the radiated particle not being a gluon, the particle which is radiated off: $\eta$ = ghost and q = light quark.}
\label{Tab:DivergentCases}
\end{table*}

As mentioned earlier, the non-trivial color structure is essential to the structure of infrared divergences. The diagrams proportional
to $C_F$ lead only to soft divergences, where one cutoff is sufficient. 
The diagrams with the color factor $C_A$, on the other hand, 
give rise to both the soft and the collinear divergences, and a two-cutoff treatment is necessary. 
The last class of diagrams is the one 
proportional to $T_f$, which is only collinear divergent. The decomposition in the different color classes simplifies the analytical and numerical cross-check of expected cancellations. 
In the following, we address each class separately.
 
% -----------------------------------------------------------------------------
\subsubsection{Soft limit}
% -----------------------------------------------------------------------------

In the soft limit the eikonal approximation allows to factorize the squared amplitudes of the diagrams containing the infrared 
divergence according to
\begin{align}
	& \!\!\!\!\! |\mathcal{M}|^2_{\mathrm{S}} ~=~ |\mathcal{M}_0|^2  \label{Eq:AnsatzSoft} \\
	& \times \left\{(-g_s^2 C_F) \left[ \frac{p_2^2}{(p_2. k)^2} +  \frac{p_3^2}{(p_3.k)^2} - \frac{2p_2.p_3}{(p_2.k)(k.p_3)} 
	\right]\right. \nonumber\\ 	
	& ~~~~~~ + \left.\frac{(-g_s^2 C_A)}{2}\left[ \frac{2p_2. p_4}{(p_2.k)(p_4.k)} +\frac{2p_3.p_4}{(p_3.k)(p_4.k)} \right]
	\right\}\,,\nonumber 
\end{align}
with $g_s$ being the strong coupling constant and $C_A =  3$ together with $C_F =  4/3$ the color factors. $|\mathcal{M}_0|^2$ is the factorized $2 \to 2$ tree-level squared matrix element. We will treat the two different color classes separately as the first proportional to $C_F$ contains only soft infrared divergences and leads to single poles in $\varepsilon$ and the second class proportional to $C_A$ combines both soft and collinear divergences. In this case, the combination of infrared divergences manifests itself through the appearance of double poles in $\varepsilon$. This difference between the two color classes requires a different treatment of the phase-space integration.

In the case of pure soft divergent amplitudes the cross-section contribution of each term of Eq.\ (\ref{Eq:AnsatzSoft}) can be written in the generic form
\begin{align}
 	\left( \frac{{\rm d} \sigma}{{\rm d} \Gamma_2} \right) = &\left( \frac{{\rm d} \sigma}{{\rm d} \Gamma_2} \right)_{\!\! 0} \frac{\, g_s^2  \mu^{4-D}}{8 \pi^3} C_F  \label{Eq:radiationterms} \\
 	& \times \int_{|\vec{k}| \leq \delta_s} \frac{{\rm d}^{D-1}k}{(2 \pi)^{D-4}} \frac{1}{2 \omega}
 \frac{a. b}{(a. k)(k. b)} , \nonumber 
\end{align}
where the integration over the $D$-dimensional momentum $k$ of the radiated particle is performed over energies up to the soft cutoff $\delta_s$. Further, $\mu$ stands for the chosen renormalization scale and ${\rm d}\Gamma_2$ denotes the two-body phase space of the tree-level final-state partons with momenta $p_3$ and $p_4$. 

In order to isolate the soft infrared divergence, we reduce the phase-space element and split off the angular part ${\rm d}\Omega_{D-2}$ which does not contain any divergence in this case
\begin{align}
	\int \frac{{\rm d}^{D-1}k}{(2\pi)^{D-1} 2\omega} = \int \frac{{\rm d}|\vec{k}|\,\, |\vec{k}|^{D-2}}{(2\pi)^{D-1} 2|\vec{k}| } 
	\int {\rm d} \Omega_{D-2}\,.
	\label{Eq:Ddimensionintegration}
\end{align}
Taking into account the $k^2$ in the denominator which arises from calculating the two scalar products in the denominator of Eq.\ (\ref{Eq:radiationterms}), the single pole can be isolated as
\begin{align}
	\int_0^{\delta_s}{\rm d}k\,\, k^{D-5} = \frac{\delta_s^{(D-4)}}{D-4} = -\frac{\delta_s^{(D-4)}}{2 \varepsilon}.
	\label{Eq:softdivergence}
\end{align}
When performing the integration over the remaining phase space, two different cases can occur: Both momenta in the scalar products of Eq.\ (\ref{Eq:AnsatzSoft}) are equal ($I_{a^2}$) or different ($I_{ab}$). The corresponding integrals can be found in App.\ \ref{App:cutoff_soft}.\\

The collinear divergences in the second and more complicated contribution to $|\mathcal{M}|^2_{\mathrm{S}}$ require additional care when integrating over the angular part of the $D$-dimensional momentum of the additional gluon. Therefore, the angular part of the integral is rewritten as \cite{HarrisOwens}:
\begin{align}
&\int \frac{{\rm d}^{D-1}k}{(2\pi)^{D-1} 2\omega} =\label{Eq:Ddimsoftcoll}\\
&\int \frac{{\rm d} k\, k^{D-2}}{(2\pi)^{D-1} 2 k} \int {\rm d}\theta_1 \sin^{D-3}\theta_1\, {\rm d}\theta_2 \sin^{D-4}\theta_2  \, {\rm d}\Omega_{D-4}. \nonumber
\end{align}
The differential cross section for the second part of $|\mathcal{M}|^2_{\mathrm{S}}$ can be re-formulated
\begin{align}
 \left( \frac{{\rm d} \sigma}{{\rm d} \Gamma_2} \right)& = -\left( \frac{{\rm d} \sigma}{{\rm d} \Gamma_2} \right)_{\!0} \frac{\, g_s^2 C_A }{16 \pi^3 } \frac{ \mu^{4-D} \pi^{\varepsilon}\,\, \Gamma(1 -\varepsilon)}{\Gamma(1 - 2 \varepsilon)} \frac{1}{s} \nonumber \\ 
 & \!\!\!\!\!\!\!\!\!\!\!\!   \times\int_0^{\delta_s}{\rm d} k\, k^{D-5} \int_0^\pi {\rm d}\theta_1 \sin^{D-3}\theta_1 \int_0^\pi {\rm d}\theta_2 \sin^{D-4}\theta_2 \nonumber \\
 & \!\!\!\!\!\!\!\!\!\!\!\!   \times  \left[ -\frac{4 (m_2^2 - t)}{x_{p_2k}\,x_{p_4k}} - \frac{2 (s - m_3^2)}{x_{p_3k}}  - \frac{2 (s - m_3^2)}{x_{p_4k}}\right],
\label{Eq:radiationtermssoftcolltrick}
\end{align}
where we use the abbreviations $x_{p_2k}, x_{p_3k}$, and $x_{p_4k}$ defined as follows:
\begin{widetext}
\begin{align}
(p_2. k) &=\frac{k \sqrt{s}}{2} \left( \frac{2 E_2}{\sqrt{s}} + \beta_1 \cos \theta_1  \right) \equiv \frac{k \sqrt{s}}{2} x_{p_2k} ,\\
(p_3. k) &= \frac{k \sqrt{s}}{2} \left( \frac{2 E_3}{\sqrt{s}} - \beta_2 \sin \theta \sin \theta_1 \cos \theta_2 - \beta_2 \cos \theta \cos \theta_1  \right) 
 \equiv \frac{k \sqrt{s}}{2} x_{p_3k} , \\
(p_4. k) &=  \frac{k \sqrt{s}}{2} \left( \frac{2 E_4}{\sqrt{s}} + \beta_2 \sin \theta \sin \theta_1 \cos \theta_2 + \beta_2 \cos \theta \cos \theta_1  \right) \equiv \frac{k \sqrt{s}}{2} x_{p_4k}.
 \label{Eq:scalarproductssoftcoll}
\end{align}
\end{widetext}
The detailed expressions for the energies $E_{2,3,4}$, as well as for $\beta_{1,2}$ can be found in App.\ \ref{App:cutoff_momenta}. For each of the three terms of Eq.\ (\ref{Eq:radiationtermssoftcolltrick}), the integration over $\theta_1$ and $\theta_2$ can be cast in the following form
\begin{align}
	I^{(l,m)}_\varepsilon = \int_0^\pi & {\rm d}\theta_1 \sin^{1 - 2\varepsilon}\theta_1\,\,\int_0^\pi {\rm d}\theta_2 \sin^{-2\varepsilon}\theta_2\nonumber\\
	& \!\!\!\!\!\! \times \frac{(a + b \cos \theta_1)^{-l}}{(A + B \cos \theta_1 + C \sin \theta_1 \cos \theta_2)^{m}}.
\label{Eq:genericsoftcollint}
\end{align}
Similar to the purely soft case, the integration over the momentum in Eq.\ (\ref{Eq:Ddimsoftcoll}), results in a soft divergence. The additional collinear divergence comes in through the integrals $I^{(l,m)}_\varepsilon$ which have been already studied for different cases of the occurring parameters in the literature, e.g. in Refs.\ \cite{vanNeerven1986,Beena1989,Smith1989,Harris1995,Bojak2000,Berger:2000iu}. 

After rearranging the relevant parts of Eq.\ (\ref{Eq:AnsatzSoft}) in order to achieve integrals of the form of Eq.\ (\ref{Eq:genericsoftcollint}) and taking into account Eq.\ (\ref{Eq:softdivergence}), we finally obtain the soft-collinear contribution
\begin{align}
 & \left( \frac{{\rm d} \sigma}{{\rm d} \Gamma_2} \right) =  \left( \frac{{\rm d} \sigma}{{\rm d} \Gamma_2} \right)_{\!0} \frac{\, g_s^2 C_A}{16 \pi^3 s} \frac{\pi^{\varepsilon}\,\, \Gamma(1 -\varepsilon)}{\Gamma(1 - 2 \varepsilon)}  \left(- \frac{1}{2 \varepsilon} \right) \left( \frac{\mu^2}{\delta_s^2} \right)^{\!\varepsilon} ~~~~ \nonumber \\
 &~~~ \times \! \left[4 (m_2^2 -t) I_\varepsilon^{(1,1)}(x_{p_2k}\,x_{p_4k}) + 2 (s - m_3^2) I_\varepsilon^{(0,1)}(x_{p_3k}) \right.\nonumber \\
 & ~~~~~~~ + \left.2 (s - m_3^2) I_\varepsilon^{(1,1)}(x_{p_4k})\right],
\end{align}
where we identify the cases $I_\varepsilon^{(0,1)}$ and $I_\varepsilon^{(1,1)}$ of the general integral given in Eq.\  (\ref{Eq:genericsoftcollint}). 

Whereas $I_\varepsilon^{(0,1)}$ results in a finite contribution, $I_\varepsilon^{(1,1)}$ gives rise to the collinear divergence. Combined with the soft divergence from the integration over the momentum, it leads to a double pole for the soft-collinear diagrams. Further details can be found in App.\ \ref{App:cutoff_softcoll}.

Thus, we have isolated all occurring soft and soft-collinear divergences of the real emission diagrams and treated them according to their color structure. The contribution proportional to $C_F$, given in Eq.\ (\ref{Eq:AnsatzSoft}), contains only single poles which are completely cancelled by the virtual counterpart without any other contribution. The contribution proportional to $C_A$ in Eq.\ (\ref{Eq:AnsatzSoft}) leads to double and single poles and only the double poles are cancelled directly by adding virtual contributions. The single poles need to be combined with other single poles from the same color class in the hard-collinear limit and only their sum cancels with the poles of the corresponding virtual contributions.  

% -----------------------------------------------------------------------------
\subsubsection{Hard-collinear limit}\label{Sec:hardcoll}
% -----------------------------------------------------------------------------

In the following, we discuss the treatment of the collinear divergences in the hard-collinear part of the phase space. The parts
of the amplitude which are not collinear divergent are not subject to the treatment described here in this section.
Collinear divergences occur, when the momentum $p_5$ of the emitted massless particle becomes collinear to the momentum $p_4$ of the massless emitter particle.\footnote{Note that in this part the momentum of the additional gluon is denoted by $p_5$ in contrast to $k$ in the previous section.} In the collinear limit, we can regard these two momenta as a single effective momentum $p_{45} = p_4 + p_5$. 
Further details on the definition of the momenta can be found in App.\ \ref{App:cutoff_momenta}.

Due to the factorization theorem \cite{Collins1985, Bodwin1984}, the squared matrix element of the $2 \rightarrow 3$ processes in the collinear limit can be described as the $2 \to 2$ matrix element multiplied by an appropriate splitting kernel
\begin{align}
	\overline{\sum} &|M_{1 + 2 \rightarrow 3 + 4 + 5}|^2 \label{Eq:leadingpoleapprox} \\
	& ~~~~ \simeq \overline{\sum} 
	|M_{1 + 2 \rightarrow 3 + 4^\prime}|^2 P_{44^\prime}(z,\varepsilon) g_s^2 \mu^{2 \varepsilon} 		\frac{2}{s_{45}},
	\nonumber
\end{align}
where $s_{45}=2~p_4.p_5$ describes the collinearity and $P_{44^{\prime}}$ the corresponding Altarelli-Parisi splitting kernels \cite{AltarelliParisi} as given in App.\ \ref{App:cutoff_hardcoll}. Thus, in the hard-collinear limit ($E_{4,5} > \delta_s$ and $0 \leq s_{45} \leq \delta_c$), the differential cross section can be written as
\begin{align}
 	\left( \frac{{\rm d} \sigma}{{\rm d} \Gamma_2} \right) = &~\left( \frac{{\rm d} \sigma}{{\rm d} \Gamma_2} \right)_0  		\frac{g_s^2 (4 \pi \mu^2)^{\varepsilon}\,\, \Gamma(1 -\varepsilon)}{8 \pi^2\Gamma(1 - 2 		\varepsilon)}  
		\label{Eq:23phasespacehardcoll3}\\
	 & \times \int_0^{\delta_c} \frac{{\rm d} s_{45}}{s_{45}^{\varepsilon + 1}} \int {\rm d}z 		\frac{P_{44^\prime}(z,\varepsilon)}{(z(1-z))^\varepsilon}, \nonumber
\end{align}
where ${\rm d}\Gamma_2$ denotes the two-body phase space of the particles with momenta $p_3$ and $p_{45}$. Moreover, $z$ describes the momentum fraction of particle with momentum $p_4$ to the quasi-particle $p_{45}$. Further details on the definitions can be found in App.\ \ref{App:cutoff_momenta} and we refer to Ref.\ \cite{HarrisOwens} for all other details on the derivation.

The collinear divergence can be isolated thanks to the fact that the integral over the momentum fraction in Eq.\ (\ref{Eq:23phasespacehardcoll3}) is independent of $s_{45}$:
\begin{align}
	\int_0^{\delta_c} \frac{{\rm d} s_{45}}{s_{45}^{\varepsilon + 1}} ~=~ -\frac{1}{ \varepsilon} 
		\delta_c^{-\varepsilon}.
\end{align}
The integration bounds of the integral over the momentum fraction $z$ in Eq.\ (\ref{Eq:23phasespacehardcoll3}) have to reflect the fact that the energies of the final state 
particles in the hard-collinear region are bound from below by the soft cutoff $\delta_s$.
Therefore, the amplitudes which correspond to the process $\tilde{\chi}_1^0 \tilde{t}_1 \rightarrow t g g$ and are divergent also in
the soft limit, have to be integrated by using the following integration bounds
\begin{align}
	1 - \frac{1 - \frac{\delta_s}{\beta}}{1 - \frac{s_{45}}{s_{12}}\frac{1}{\beta}} 
	\leq z \leq 1 - \frac{1 - \frac{\delta_s}{\beta}}{1 - \frac{s_{45}}{s_{12}}\frac{1}{\beta}},
	\label{Eq:integrationboundsgg}
\end{align}
with
\begin{align}
	\beta = 1 - \frac{m_3^2}{s_{12}}.
\end{align}
Further details on the derivation can be found in App.\ \ref{App:cutoff_hardcoll}.

There is still a class of diagrams which was not yet mentioned. 
It is the one belonging to the processes $\tilde{\chi}_1^0 \tilde{t}_1 \rightarrow t q \overline{q}$ with 
light quarks in the final state. The squared matrix elements of this class contain the color factor $T_f$. 
These diagrams do not give rise to any soft divergences but are collinear divergent. The integration in this case is possible
over the whole momentum fraction phase space $0 \leq z \leq 1$. 

Performing all integrations, the differential cross section can be written as
\begin{align}
 	\left( \frac{{\rm d} \sigma}{{\rm d} \Gamma_2} \right) & = \left( \frac{{\rm d} \sigma}{{\rm d} \Gamma_2} \right)_{\!0}  	\frac{g_s^2 }{8 \pi^2} \biggr[ \left( A^{g\rightarrow gg}_0 + A^{g\rightarrow q \overline{q}}_0 \right) ~~~~~ \\
	 & \!\!\!\!\!\!\!\!\!\!\! + \left( \frac{1}{\varepsilon} + \log 4 \pi - \gamma_E + \log\mu^2 \right) \Big( A^{g\rightarrow gg}_\varepsilon + A^{g\rightarrow q \overline{q}}_\varepsilon \Big) \biggr],
	 \nonumber
\end{align}
with the corresponding form factors given by
\begin{align}
	A^{g\rightarrow q \overline{q}}_0 &= \frac{n_f}{3}\left[ \log \delta_c - \frac{5}{3} \right], \\
	A^{g\rightarrow q \overline{q}}_\varepsilon &= -\frac{n_f}{3} , \\
	A^{g\rightarrow gg}_0 &= C_A \left[\frac{67}{18} - \frac{\pi^2}{3} - \biggr( \log 2 \delta_s - \log ( \sqrt{s_{12}} - \frac{m_3^2}{\sqrt{s_{12}}}) \biggr)^{\!2} \nonumber \right. \\
	& ~~ - \left. \log \delta_c \left(\frac{11}{6} + \log 4 \delta_s^2 - 2\log ( \sqrt{s_{12}} - 		\frac{m_3^2}{\sqrt{s_{12}}} )\right) \right. \nonumber \\
	& ~~ \left. + ~2~ \mathrm{Li}_2 \!\left( \frac{\delta_c}{2~ \sqrt{s}~ \delta_s}\right) \right] ,\\
	A^{g\rightarrow gg}_\varepsilon &= C_A \left[ \frac{11}{6} + \log 4 \delta_s^2 - 2\log \left( \sqrt{s_{12}} - 		\frac{m_3^2}{\sqrt{s_{12}}} \right) \right] .
\end{align}
Further details on the derivation can be found in Ref.\ \cite{HarrisOwens}. 
A different definition of the soft cutoff $\delta_s$ in the previous section and the fact that we have considered a massive particle with momentum $p_3$ lead to a small difference in expressions when compared to Ref.\ \cite{HarrisOwens}. These differences have to be taken also into account in the $\mathcal{O}(\delta_c/\delta_s)$ terms which we have also included in our analysis.

% -----------------------------------------------------------------------------
\subsubsection{Cutoff independence}
% -----------------------------------------------------------------------------

\begin{figure*}
    \includegraphics[scale=0.4]{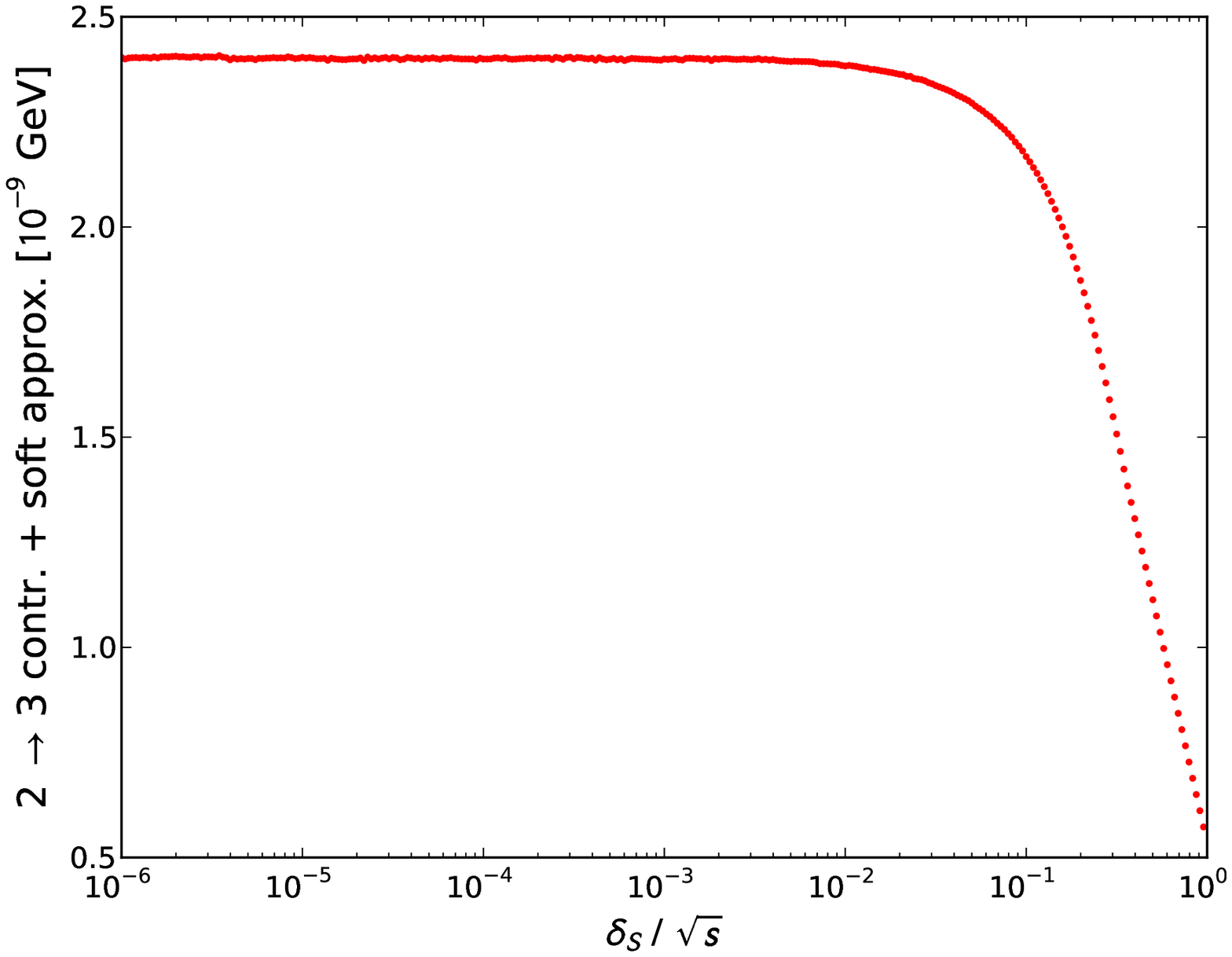}
    \includegraphics[scale=0.4]{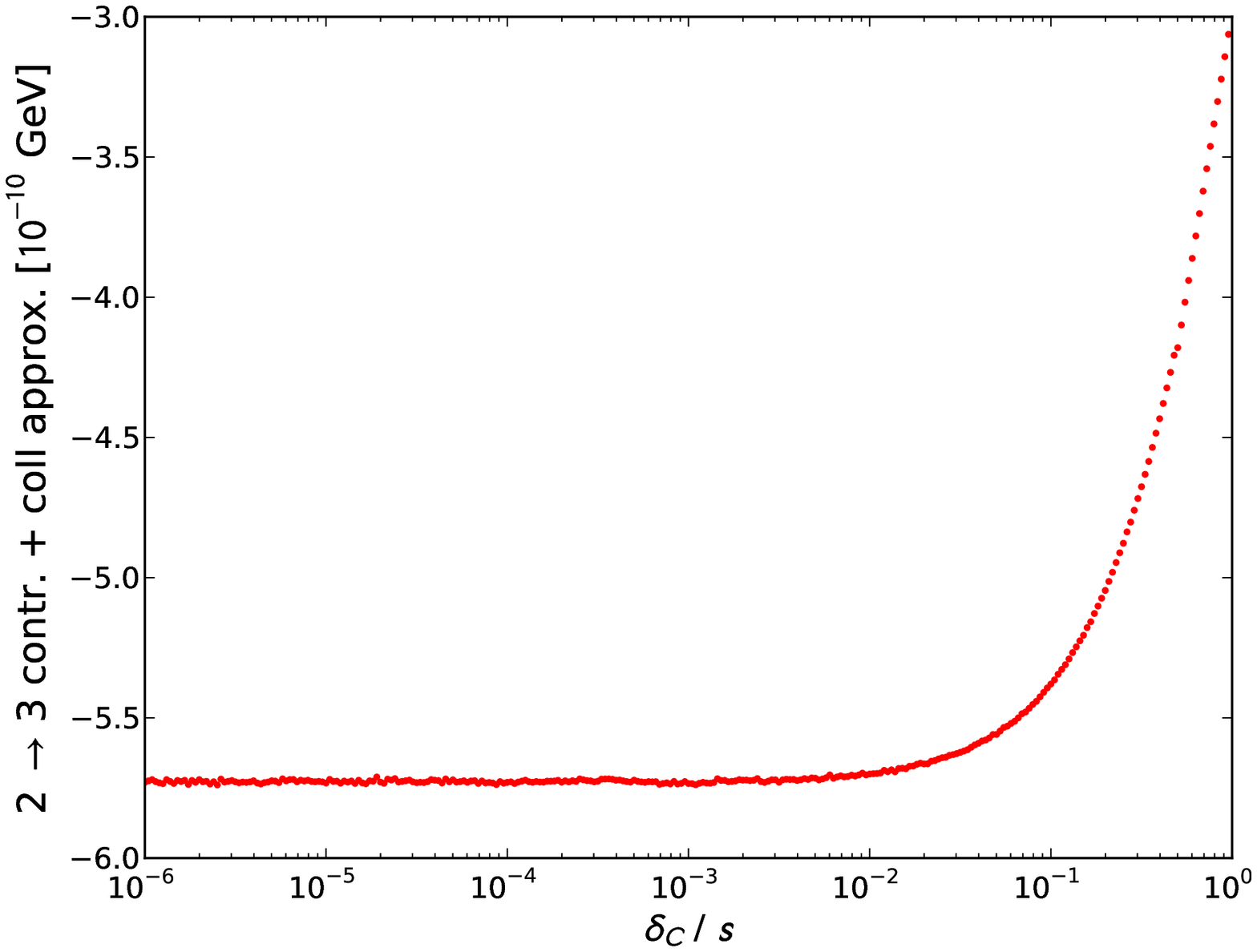}
	\includegraphics[scale=0.45]{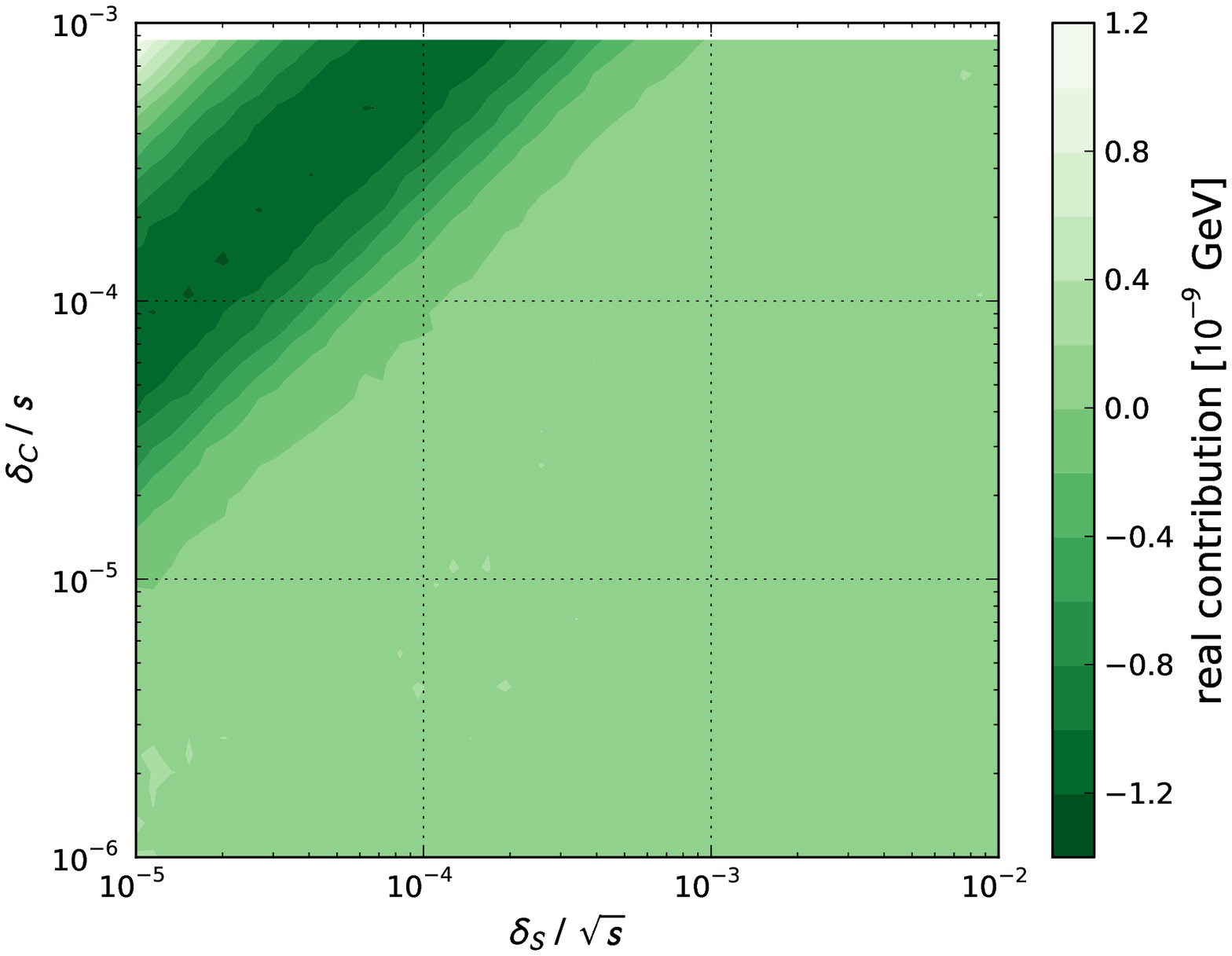}
	\caption{Plots demonstrating the cutoff independence of our NLO calculation for the case of a final state with a top quark and a gluon. Upper left: Purely soft IR-convergent contribution arising from gluon emission off quarks and squarks. Upper right: Purely collinear IR-convergent contribution arising from light quark emission. Bottom: Soft-collinear IR-convergent contribution arising from gluon emission off a gluon.} 
	\label{Fig:CutoffPlots}
\end{figure*}

The phase-space slicing method allows for the cancellation of all occurring infrared divergences over the whole phase space. In the calculation of the full $2 \rightarrow 3$ process, this method introduces a dependence on the cutoffs $\delta_s$ and $\delta_c$
\begin{equation}
	\sigma^{2\rightarrow 3}_{\rm full} = \sigma_{\rm S}(\delta_s) + \sigma_{\rm HC}(\delta_s,\delta_c) + 
	\sigma_{{\rm H}\overline{\mathrm{C}}}(\delta_s,\delta_c).
\end{equation}
The final result, however, does not depend on these cutoffs if the splitting is performed properly. The cutoff independence is therefore a powerful numerical check, especially since the different contributions as classified in Tab.\ \ref{Tab:DivergentCases} can be investigated separately. An example of such a check is shown in Fig.\ \ref{Fig:CutoffPlots}.  

First, we show two special subsets of matrix elements where the infrared divergence comes either only from the soft or only from the collinear region. It stands to reason that these subsets of matrix elements need only one cutoff for the phase-space slicing to isolate the divergence, either the soft cutoff $\delta_s$ or the collinear cutoff $\delta_c$.

For the process with a gluon in the final state, the upper-left plot of Fig.\ \ref{Fig:CutoffPlots} shows the behavior of the purely soft contributions to the cross section. The relevant squared amplitudes are given in the second row of Tab.\ \ref{Tab:DivergentCases}. They include gluon emission from the initial stop, the propagator top or stop, respectively, and the final state top. The graph shows the sum of the corresponding $2 \rightarrow 3$ processes and their soft gluon approximation. If the cutoff independence is fulfilled, the sum of both should stay constant when varying the soft cutoff parameter while being independent from the collinear cutoff altogether. In the upper-right subfigure of Fig.\ \ref{Fig:CutoffPlots} a similar plot for the purely hard-collinear light quark emission is shown. Here the corresponding $2 \rightarrow 3$ processes and their collinear approximation are added up. As there are no soft-collinear contributions in this case, it can be studied independently from the soft cutoff. The collinear approximation is well valid up to roughly $\delta_c/s = 2 \cdot 10^{-3}$.

Generally, for too small values for the cutoffs problems can occur regarding the cutoff independence. In this case, the full 
$2 \rightarrow 3$ matrix element would be integrated already over a part of the phase space where the divergence resides, which
renders the numerical integration unreliable. Similar problems would arise for too large cutoffs, where we reach a region where the soft (or collinear) limit does not hold any more and the approximation breaks down. Therefore, we have checked for cutoff independence in a limited interval, e.g., for the soft cutoff $\delta_s / \sqrt{s} \in (10^{-5},10^{-2})$ or the collinear cutoff $\delta_c/s \in (10^{-6}, 10^{-3})$. As can be seen in Fig.\ \ref{Fig:CutoffPlots}, the result is indeed independent of the cutoff over the given intervals.

As discussed in Sec.\ \ref{Sec:hardcoll}, the matrix elements which include the $g \rightarrow g g$ splitting are divergent both
in the soft and in the collinear limit, which is clearly manifested by the appearance of double poles. As a consequence, the cutoff behavior can be studied only in dependence of both, the soft and collinear cutoff. This is shown in the lower plot of Fig.\ \ref{Fig:CutoffPlots}. Here, a broad plateau is visible, which demonstrates that also in the soft-collinear case the calculation is independent from both cutoffs over a large region of phase space. For our study we have chosen the cutoffs to be $\delta_s / \sqrt{s} = 3.0 \cdot 10^{-4}$  and $\delta_c / s = 3.0 \cdot 10^{-5}$. We have verified that the chosen value for the soft cutoff is also appropriate for the other possible coannihilation final states, e.g., for the $t h^0$ final state.

% =============================================================================
%!TEX root=paper.tex
\begin{table*}
		\begin{center}
		\begin{tabular}{|ccccccccccc|}
			\hline 
			    ~$\tan\beta$~ & ~~~$\mu$~~~ & ~~~$m_A$~~~ & ~~~$M_1$~~~ & ~~~$M_2$~~~ & ~~~$M_3$~~~ & ~~$M_{\tilde{q}_{1,2}}$~~ & ~~~$M_{\tilde{q}_3}$~~~ & ~~~$M_{\tilde{u}_3}$~~~ & ~~~$M_{\tilde{\ell}}$~~~ & ~~~$T_t$~~~  \\
			\hline
			   5.8 & 2925.8 & 948.8 & 335.0 & 1954.1 & 1945.6 & 3215.1 & 1578.0 & 609.2 & 3263.9 & 2704.1 \\  
			\hline
		\end{tabular}
		\end{center}
		\caption{Parameters in the pMSSM defining the example scenario in the pMSSM. All quantities except $\tan\beta$ are given in GeV.}	
		\label{Tab:Scenarios}
\end{table*}

\begin{table}[h!]
		\begin{center}
		\begin{tabular}{|c|c|}
			\hline 
			$m_{\tilde{\chi}^0_1}$ & 338.3 GeV \\
			$m_{\tilde{t}_1}$ & 375.6 GeV \\
			$m_{h^0}$ & 122.0 GeV \\
			\hline
			$\Omega_{\tilde{\chi}_1^0}h^2$ & 0.1136 \\
			${\rm BR}(b\to s\gamma)$ & $3.25 \cdot 10^{-4}$ \\
			\hline
		\end{tabular}
		\end{center}
		\caption{Physical neutralino, stop, and Higgs masses as well as neutralino relic density and the inclusive branching ratio of the decay $b\to s\gamma$.}	
		\label{Tab:Masses}
\end{table}

\section{Results and discussion \label{Sec:Results}}
% =============================================================================

In the following, we discuss the impact of the radiative corrections on the (co)annihilation cross section and the neutralino relic density on the basis of an example scenario as defined in Tab.\ \ref{Tab:Scenarios}. For the sake of generality, we have chosen to work in the phenomenological MSSM (pMSSM) with eleven free parameters. In this setup, the Higgs sector is parametrized by the ratio of the vacuum expectation values of the two Higgs doublets, $\tan\beta$, the off-diagonal Higgsino mass parameter $\mu$, and the pole mass of the pseudoscalar Higgs boson, $m_A$. The bino, wino, and gluino mass parameters $M_1$, $M_2$, and $M_3$ are chosen to be independent, which allows the most general situation at the level of the decomposition of the neutralino dark matter candidate. The masses of left- and right-handed squarks of the first and second generation are parametrised by a common mass parameter $M^2_{\tilde{q}_{1,2}}$. For the third generation of up-type squarks, we have two parameters $M^2_{\tilde{q}_3}$, corresponding to left-handed stops and sbottoms as well as right-handed sbottoms, and $M^2_{\tilde{u}_3}$ for the right-handed stops. The trilinear coupling parameter for the stops is $T_t=A_t Y_t$, while the remaining trilinear couplings are set to zero. Finally, the slepton sector being less relevant in our study, we restrict our analysis to a common mass parameter $M^2_{\tilde{\ell}}$ for all left- and right-handed sleptons and sneutrinos.

The corresponding mass spectrum is obtained using the public spectrum generator {\tt SPheno 3.2.3} \cite{SPheno}. We show the most relevant masses like the mass of the lightest neutralino, the lightest stop, and the light $CP$-even Higgs-boson in Tab.\ \ref{Tab:Masses}. The neutralino relic density given in Tab.\ \ref{Tab:Masses} has been obtained by using the standard {\tt micrOMEGAs 2.4.1} calculation. Finally, the value of the inclusive branching ratio of the decay $b\to s\gamma$ as obtained by {\tt SPheno} is also indicated in Tab.\ \ref{Tab:Masses}.

% -----------------------------------------------------------------------------
\subsection{Phenomenology}
% -----------------------------------------------------------------------------

\begin{figure*}
	\begin{center}
		\includegraphics[scale=0.4]{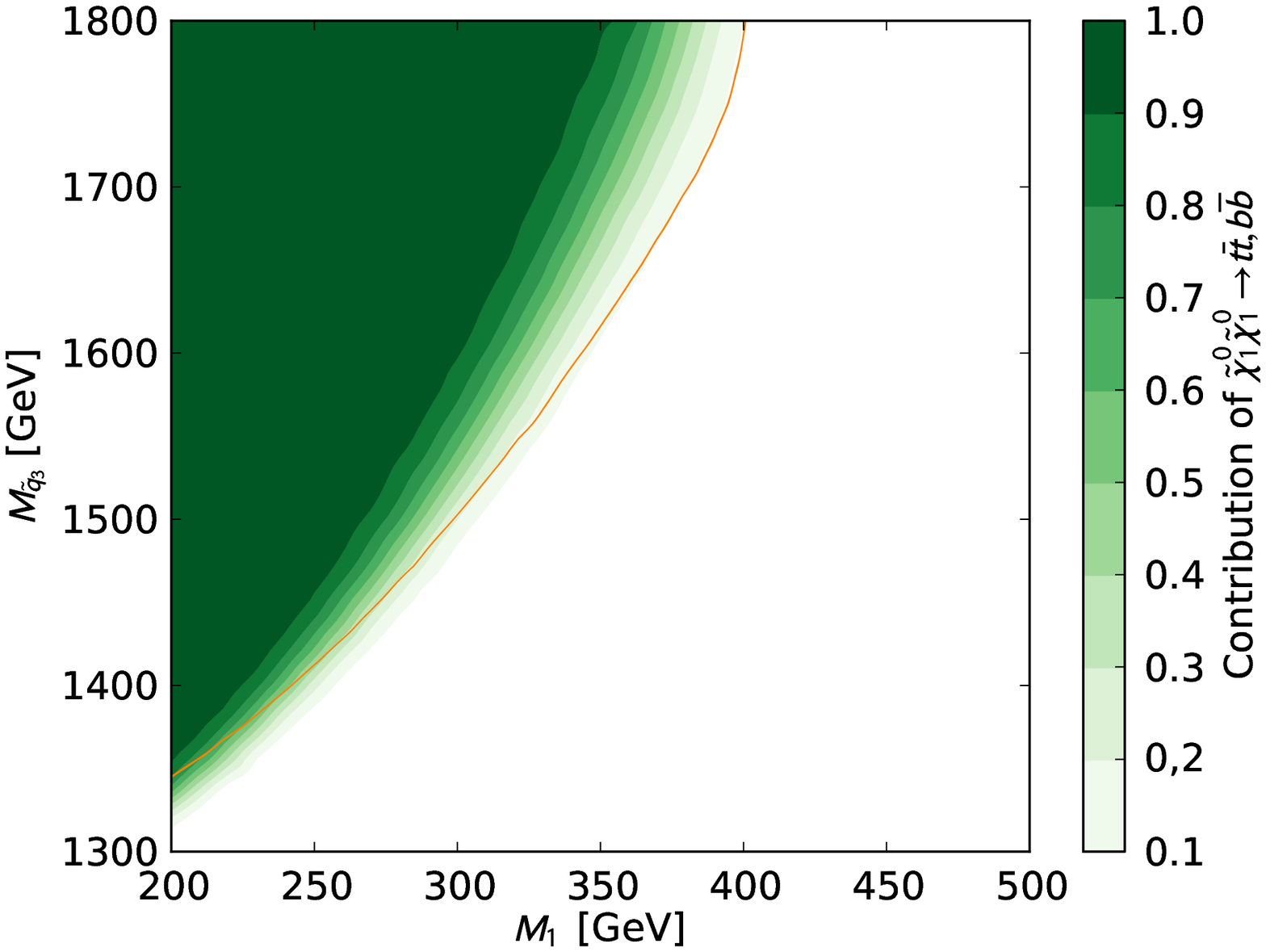}
		\includegraphics[scale=0.4]{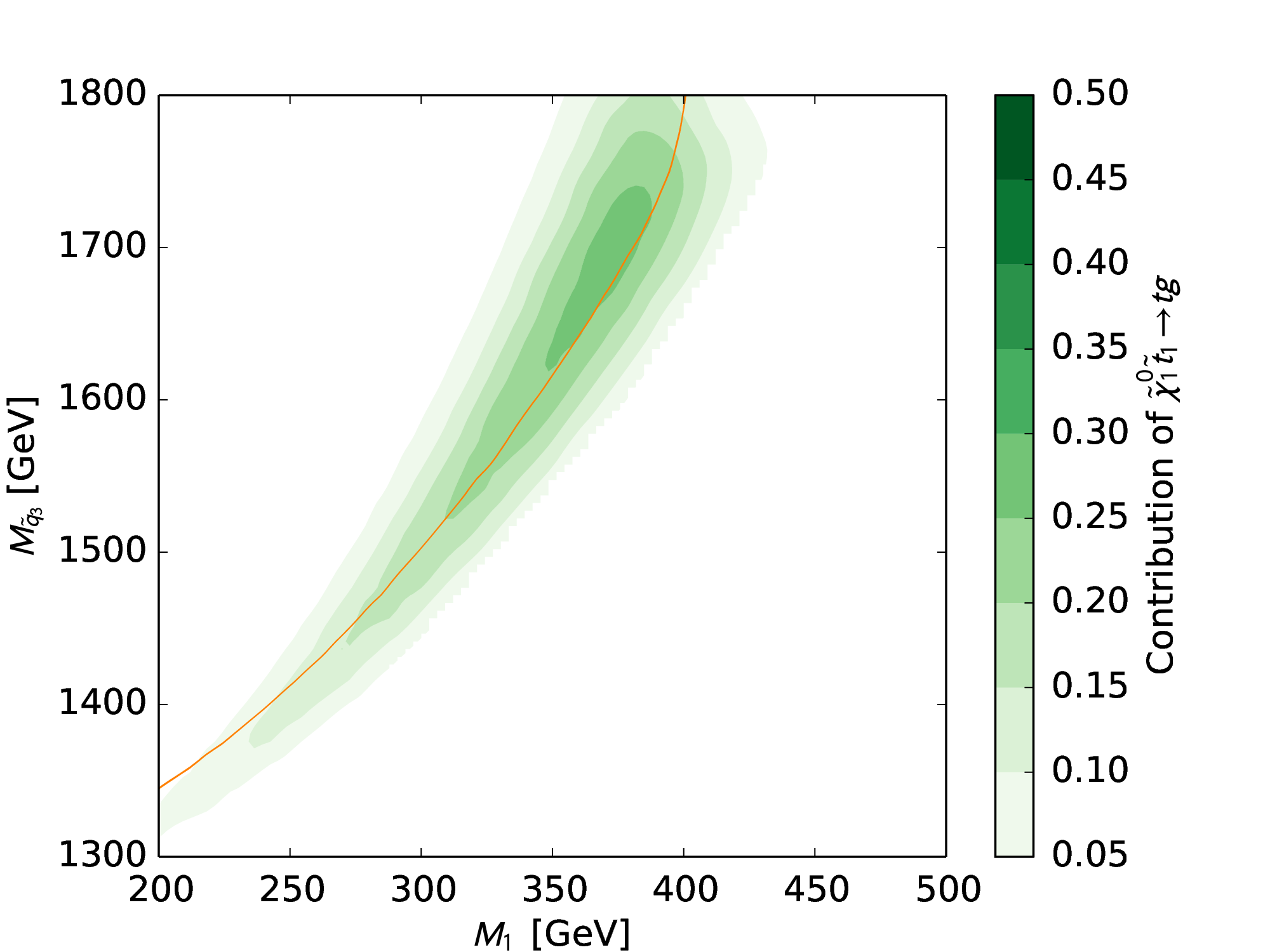}
		\includegraphics[scale=0.4]{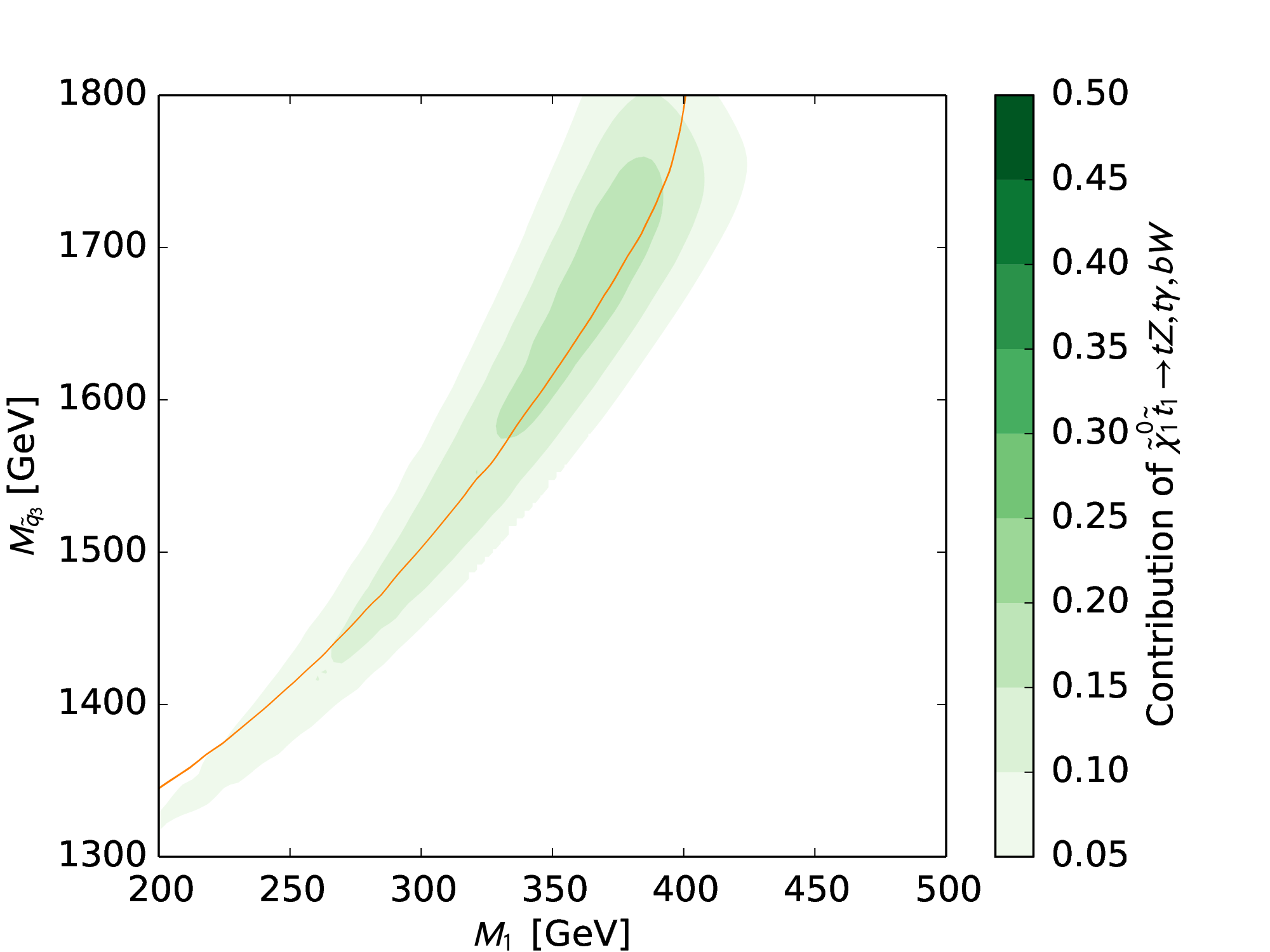}
		\includegraphics[scale=0.4]{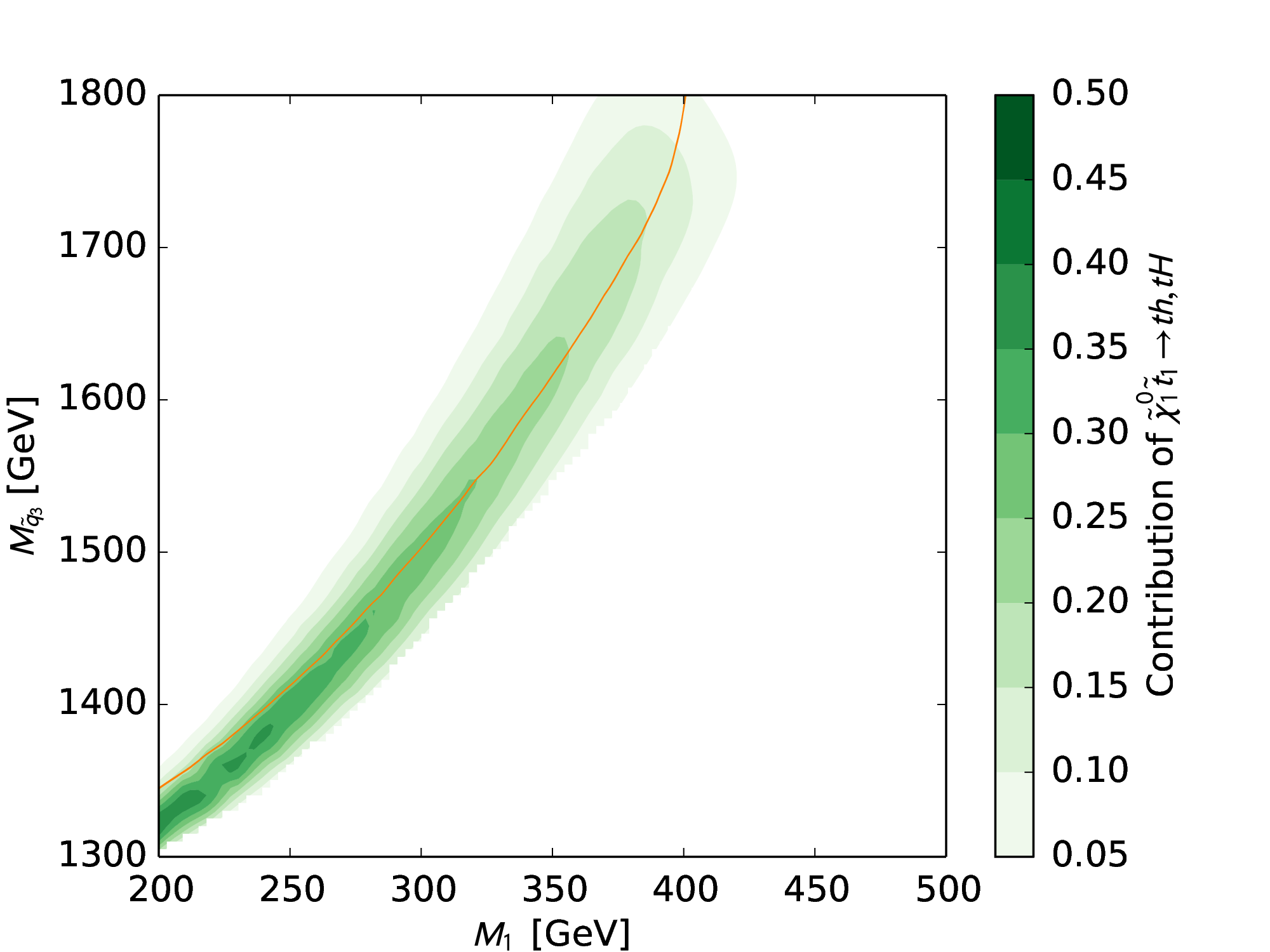}
	\end{center}
	\caption{Two-dimensional scan in $M_1$ and $M_{\tilde{q}_3}$ showing the relative contribution of different processes to the total (co)annihilation cross section $\sigma_{\mathrm {ann}}$. The orange band indicates the region of parameter space which is (at 1$\sigma$ confidence level) favored by the Planck measurement corresponding to Eq.\ (\ref{Eq:Planck}).}
	\label{Fig:BinoSquarkplane}
\end{figure*}

Before studying in detail the impact of the loop corrections on the cross section and on the neutralino relic density, we want to 
discuss the phenomenology of the chosen scenario. In Fig.~\ref{Fig:BinoSquarkplane}, we show the cosmologically favored region 
(see Eq.\ (\ref{Eq:Planck})) in the $M_1$--$M_{\tilde{q}_3}$ parameter plane together with the four leading
contributions to the total (co)annihilation cross section $\sigma_{\mathrm {ann}}$. The other remaining free parameters are set as indicated in Tab.\ \ref{Tab:Scenarios}. The region in parameter space where the relic density is compatible with the measured value by the Planck satellite within one sigma, is denoted by an orange band.

The Planck preferred region in the case of our scenario follows an approximate straight line of constant mass difference between the lightest neutralino and the lightest scalar top quark. Comparing the plots in Fig.\ \ref{Fig:BinoSquarkplane}, we see that the preferred region runs through areas where different (co)annihilation processes dominate.

For larger values of both $M_1$ and $M_{\tilde{q}_3}$, the coannihilation into a final state with a vector boson dominates. Out of all possible vector bosons, the gluon gives the largest contribution because of the strong interaction of the gluon with the top quark. This contribution can be as large as 30\%, whereas the other vector contributions together are typically only half as important -- up to 10-20\%.

In the opposite corner of the preferred region, where the importance of the final state with a vector boson diminishes, the neutralino and the scalar top quark coannihilate predominantly into a Higgs boson and a quark (see the lower-right plot in Fig.\ \ref{Fig:BinoSquarkplane}). This process was analyzed in Ref.\ \cite{DMNLO_Stop1},  where it was shown that the importance of this process is connected to the large trilinear coupling in the $t$-channel exchange diagram. Our scenario features such a large trilinear coupling as a means to satisfy the Higgs mass constraint. Therefore, it is to be expected that this channel becomes very important for smaller values of $M_{\tilde{q}_3}$, which enhances the contribution of the $t$-channel diagram in this process. The contribution of this class of processes to the relic density can be as high as 40\%.

The last important contribution is the neutralino annihilation into a top anti-top pair (see the upper-left plot in Fig.~\ref{Fig:BinoSquarkplane}). This contribution lies predominantly above the preferred region and it is more important for lighter stop masses as the driving matrix element is the scalar top quark exchange in the $t$-channel.

The parameter point of Tab.\ \ref{Tab:Scenarios} is chosen within the plane of Fig.\ \ref{Fig:BinoSquarkplane}. It features a neutralino relic density of $\Omega_{\tilde{\chi}^0_1}h^2 = 0.1136$ and a Higgs mass of $m_{h^0} = 122.0$ GeV, which both lie within theoretical and experimental uncertainties. The mass difference between the lightest stop and the lightest neutralino is 37.3 GeV, which favors their coannihilation. The corresponding relative contributions to the total neutralino annihilation cross section of Eq.\ (\ref{Eq:Sigma}) of the dominant processes are listed in Tab.\ \ref{Tab:Channels}. With the neutralino annihilation and neutralino-stop coannhilation processes summing up to $76\%$, we are able to correct a large fraction of the total coannhilation cross section. 

\begin{table}
		\begin{center}
		\begin{tabular}{|c|c|}
			\hline
				$\tilde{\chi}_1^0 \tilde{t}_1 \to t g$ 		& 23\%  \\
			    $\tilde{\chi}_1^0 \tilde{t}_1 \to t h^0$ 	& 23\%   \\
			    $\tilde{\chi}_1^0 \tilde{t}_1 \to t Z^0$ 	& 5\%     \\
			    $\tilde{\chi}_1^0 \tilde{t}_1 \to b W^+$ 	& 10\%   \\
			\hline
			    $\tilde{\chi}_1^0 \tilde{\chi}_1^0 \to t \bar{t}$	    & 15\%    \\
			    \hline
			     \hline
			   $\sum_{\mathrm{corr}}$  & 76\%  \\[1mm]
			    \hline
		\end{tabular}
		\end{center}
		\caption{Dominant (co)annihilation channels contributing to $\Omega_{\tilde{\chi}_1^0}h^2$ for the example scenario of Tab.\ \ref{Tab:Scenarios}.}	
		\label{Tab:Channels}
\end{table}

% -----------------------------------------------------------------------------
\subsection{Coannihilation cross section \label{sec:crosssec}}
% -----------------------------------------------------------------------------

\begin{figure*}
	\begin{center}
        \includegraphics[scale=0.4]{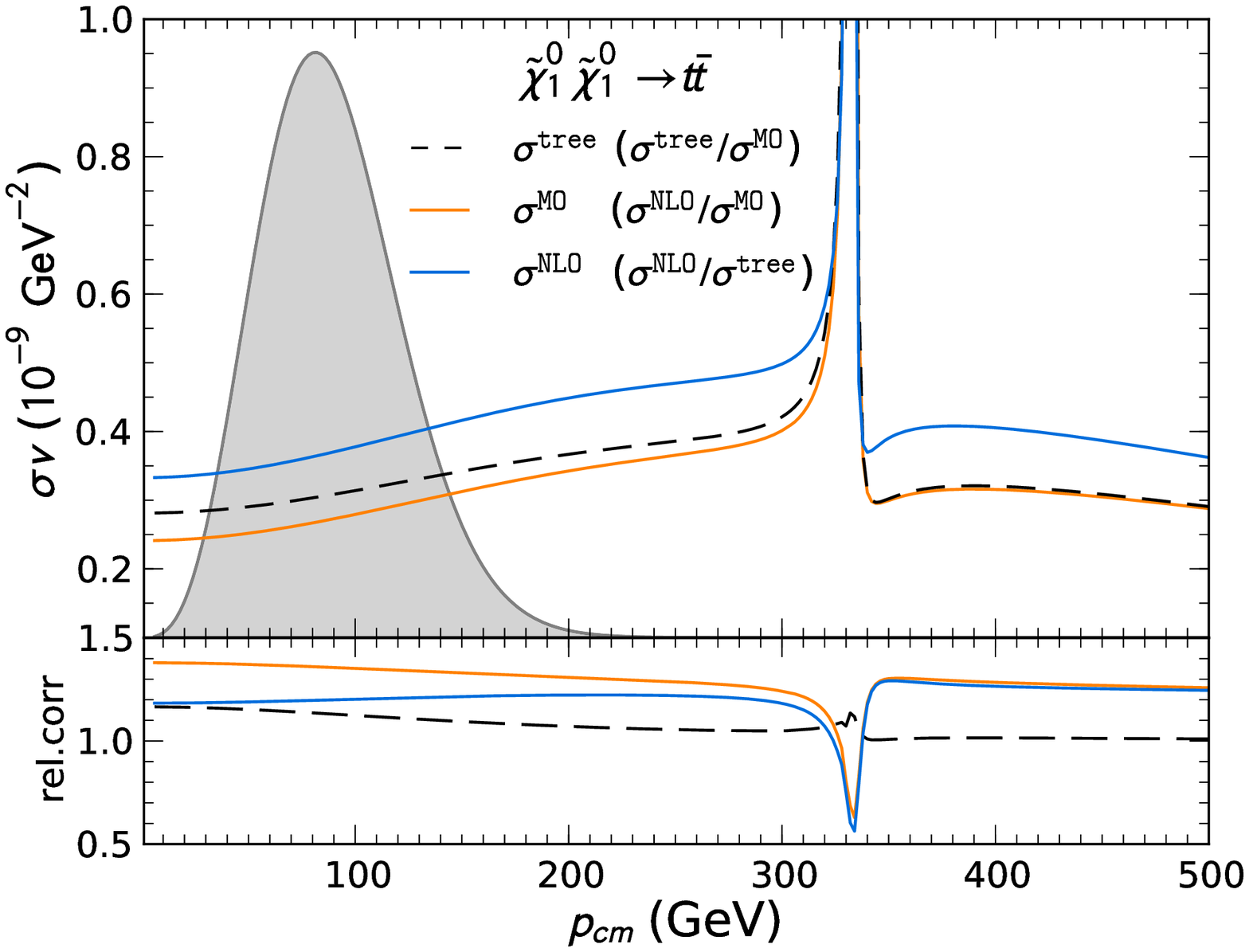}
        \includegraphics[scale=0.4]{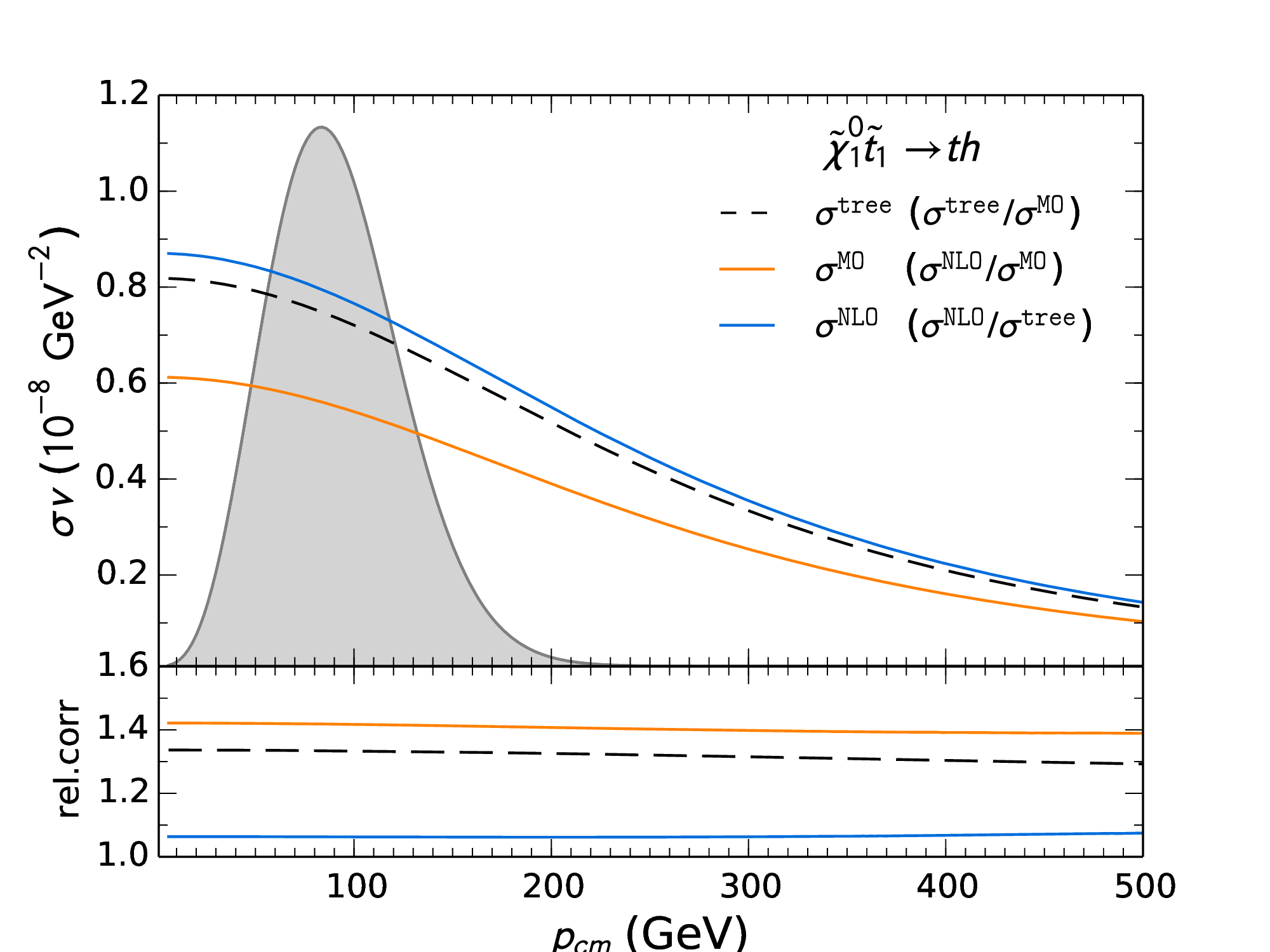}
        \includegraphics[scale=0.4]{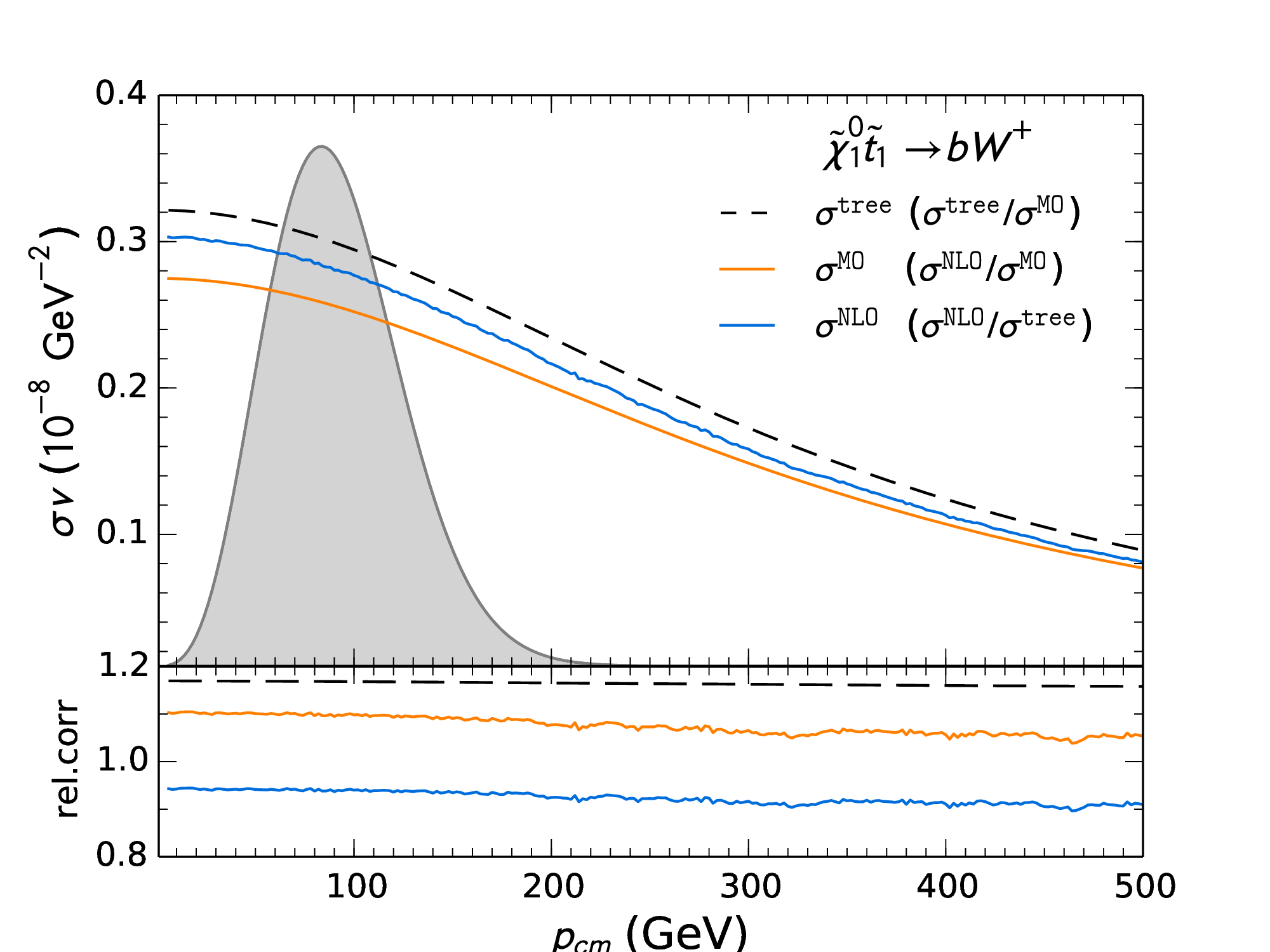}
		\includegraphics[scale=0.4]{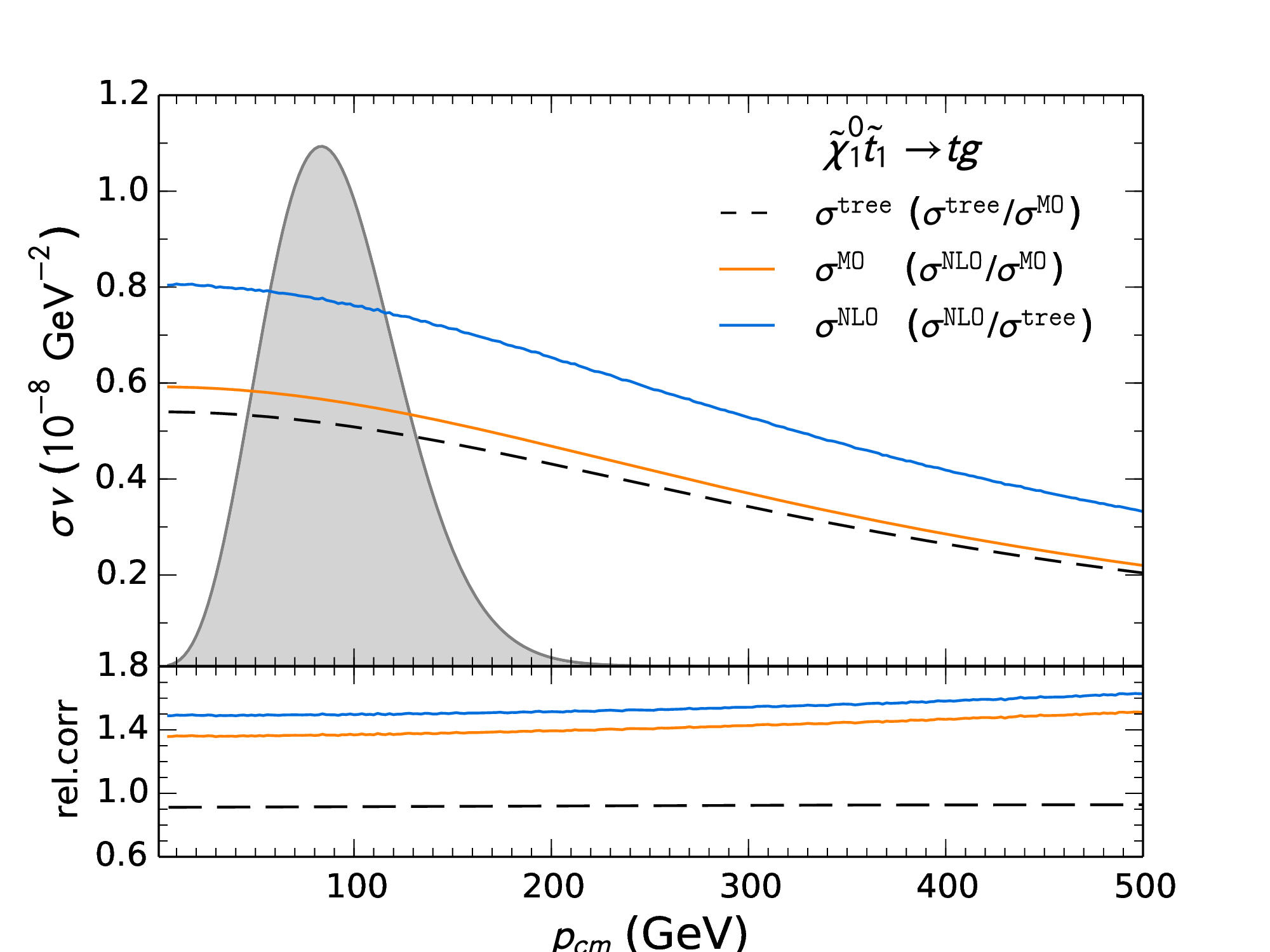}
	\end{center}
	\caption{Tree-level (black dashed line), full one-loop (blue solid line) and {\tt micrOMEGAs} (orange solid line) cross sections for the dominant (co)annihilation channels. The upper part of each plot shows the absolute value of $\sigma v$ together with the thermal velocity distribution (in arbitrary units), whereas the lower part shows the corresponding relative shift (second item in the legend).}
	\label{Fig:XSecI}
\end{figure*}

In the following, we study the impact of the one-loop corrections on the cross section of the different (co)annihilation 
sub-channels. In Fig.~\ref{Fig:XSecI}, we show the cross section of the four dominant (co)annihilation channels as a function of the centre-of-mass momentum $p_{\mathrm{cm}}$. For each channel we show our tree-level (black dashed line), the full one-loop (blue solid line), and the {\tt micrOMEGAs} (orange solid line) cross section. The grey shaded area depicts the thermal velocity distribution in arbitrary units in order to demonstrate in which region of $p_{\mathrm{cm}}$ the cross section contributes to the neutralino relic density. Furthermore, in the lower part, we show the corresponding relative shifts of the differently calculated cross sections (second item in the legend).

The upper-left plot shows the cross section for neutralino annihilation into tops. The sharp peak around $p_{\mathrm{cm}} \approx 330$~GeV corresponds to the resonances of the heavier $CP$-even and pseudoscalar Higgs bosons.

A difference between our and the {\tt micrOMEGAs} cross section at tree level can be observed. However, differences even at tree level are a general and expected feature. The first general difference is the different definition of masses and mixing matrices due to our chosen renormalization scheme. Another difference is that {\tt micrOMEGAs} provides a tree level with effective couplings. Especially, the treatment of the top quark mass gives rise to large differences. Whereas in our calculation we use the on-shell mass for the top with $m_t^{\tt OS}=172.3$ GeV, in {\tt micrOMEGAs} the ${\overline {\tt DR}}$-mass with  $m_t^{\overline {\tt DR}} = 161.6$ GeV is taken into account. This can lead to large differences in particular in diagrams containing Yukawa couplings. In case of neutralino pair-annihilation this kind of different technical treatment causes only a relatively small difference of around 10\%. If we consider further the effect of our one-loop calculation, a large shift of around 35\% with respect to the default {\tt micrOMEGAs} result is obtained. 

In the upper-right plot, a similar plot for neutralino-stop coannihilation into a top and the lightest Higgs boson is shown. In this case, the relative difference between both tree-level calculations is roughly 35\%. This is again triggered by the difference of the definition of the top mass which has a huge impact due to the Yukawa couplings in the $s$- and $t$-channel diagrams. With the $t$-channel being enhanced due to a large trilinear coupling, the effect is even larger. This leads also to a huge shift between the one-loop corrected result with respect to {\tt micrOMEGAs} of around 43\%. However, we can observe that our NLO calculation with respect to our own tree-level calculation differs only by less than 10\%, which confirms perturbativity.

In the case of an electroweak vector boson in the final state, the difference of the two tree-level calculations lies at around 15\%. This is smaller than for the Higgs boson final state, as the Yukawa couplings are less important. However, the effect is larger than for the neutralino pair annihilation, as the latter is more phase space suppressed and the $s$-channel dominated $W$ boson final state is more sensitive to the exact definition of the mixing angles. With the loop corrections being negative (in contrast to the channels discussed before) this, however, causes smaller loop corrections of about 10\%.

However, the process with the gluon in the final state shows a completely different behavior. Here, our tree-level cross section with respect to the one of {\tt micrOMEGAs} is lower by 9\%. In this case, the main reason for the difference lies in the fact that for the gluon final state the renormalized $\alpha_s$ already enters at the tree level and thus causes a difference. After accounting for the next-to-leading order contribution, the SUSY-QCD corrections lead to a positive shift of roughly 40\% over {\tt micrOMEGAs} result. This is usual for such calculations and caused by the strong loop corrections containing $\alpha_s^2$.

Altogether, the one-loop corrections account for relative corrections of the range of 10\% to 45\% with respect to the cross sections used by {\tt micrOMEGAs}. This shows the necessity to take into account these loop corrections and to further study their impact on the neutralino relic density.

% -----------------------------------------------------------------------------
\subsection{Neutralino relic density \label{sec:neutralinorelic}}
% -----------------------------------------------------------------------------

\begin{figure*}
	\begin{center}
		\includegraphics[scale=0.4]{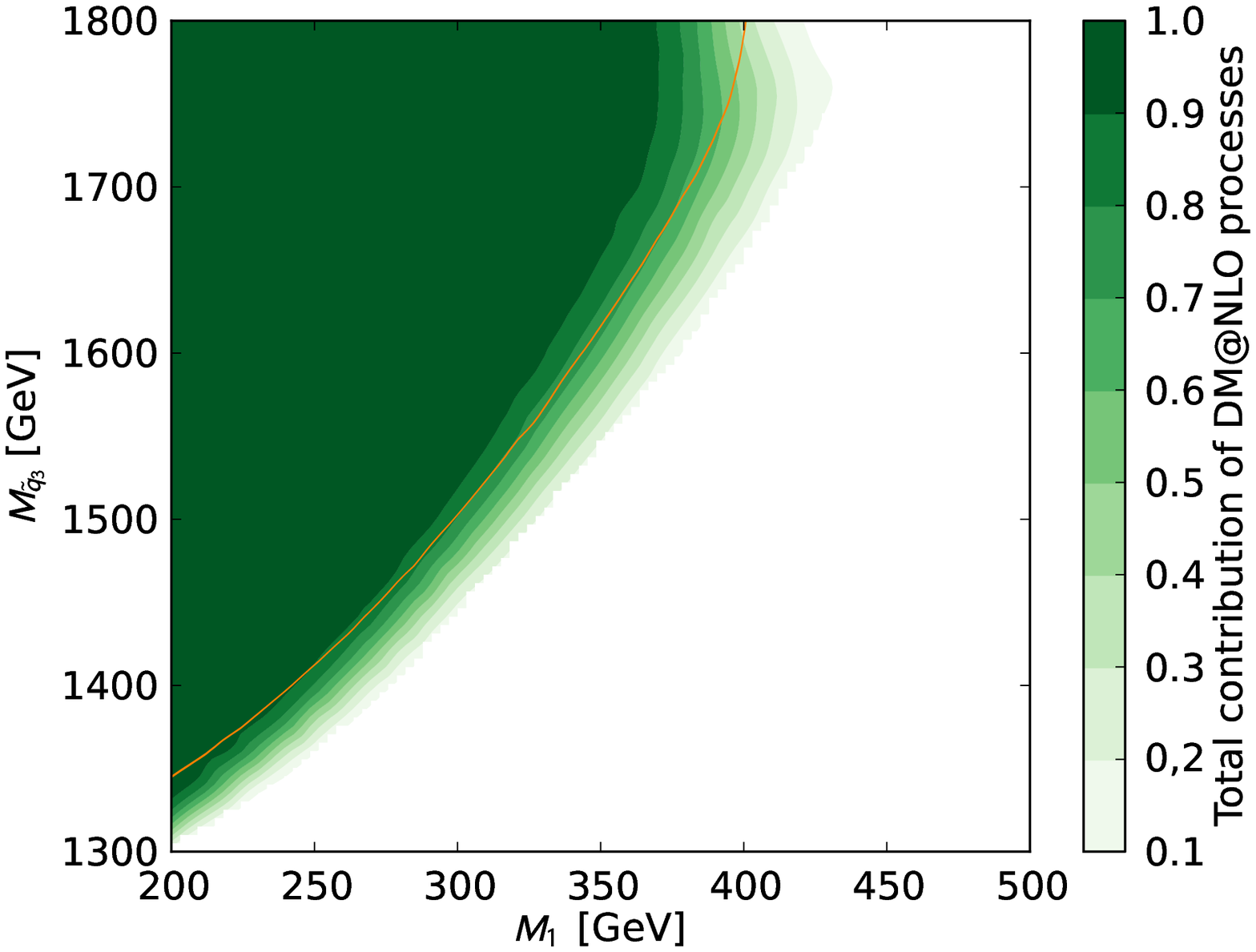}
		\includegraphics[scale=0.4]{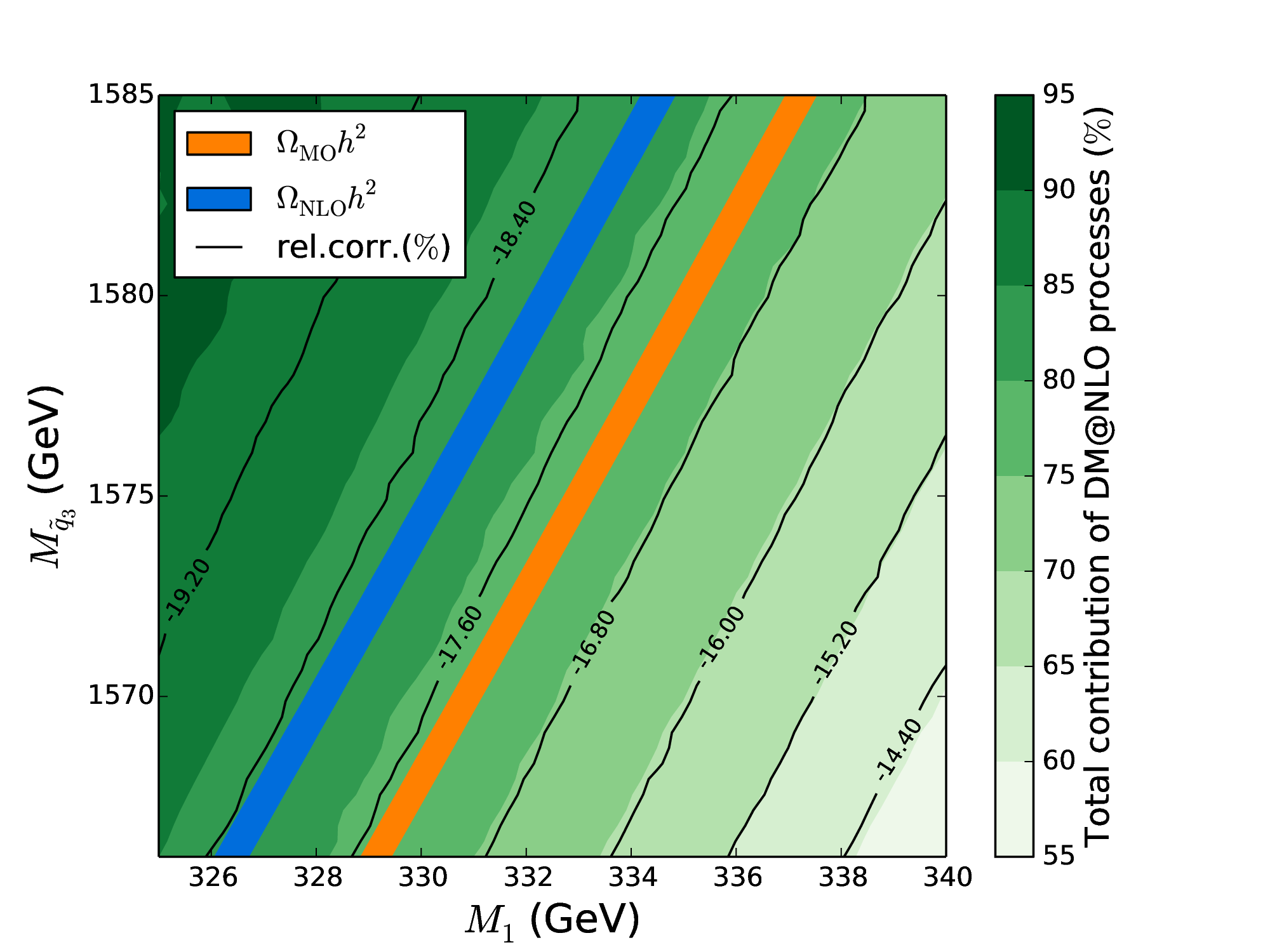}
	\end{center}
	\caption{Scan in the $M_1$--$M_{\tilde{q}_3}$ plane in the vicinity of our example scenario. On the left hand side we show in green the total contribution of corrected (co)annihilation processes. The favored region of parameter space by the one-sigma Planck results is depicted in orange. On the right hand side a zoom-in is shown. The blue band shows additionally the favored Planck one-sigma band taking into account the presented loop corrections.}
	\label{Fig:2Drelic}
\end{figure*}

For studying the impact of the next-to-leading order corrections on the neutralino relic density, we again have a look at the two-dimensional $M_1$--$M_{\tilde{q}_3}$ plane in the vicinity of our example scenario. The left plot of Fig.~\ref{Fig:2Drelic} shows in green the total contribution of processes we have corrected. As before, we use orange to highlight the parameter space compatible up to one-sigma with the Planck measurement. On the right hand side a zoom-in of this parameter plane is shown. The region favored by the Planck result is again indicated in orange color. This region is based on the out-of-the-box {\tt micrOMEGAs} calculation. In blue, we show our one-loop result. This calculation includes all SUSY-QCD corrections to neutralino annihilation into heavy quarks as well as to neutralino-stop coannhilation into all final states. We are able to correct up to 80\% of the total (co)annihilation processes of which around 65\% are coannhilation processes.

As discussed in Sec.\ \ref{sec:crosssec} the one-loop-corrected cross sections of the various channels differ from the default {\tt 
micrOMEGAs} cross section by up to 45\%. This leads to a relative correction to the {\tt micrOMEGAs} relic density of 17.5\% 
(shown as black solid lines in Fig.\ \ref{Fig:2Drelic}). This correction causes a clear shift of the Planck favored one-sigma band in the parameter space. The separation of the bands shows that the calculated loop corrections are important and have to be taken into account as their impact can be larger than the corresponding experimental uncertainty.

\begin{figure*}
	\begin{center}
		\includegraphics[scale=0.435]{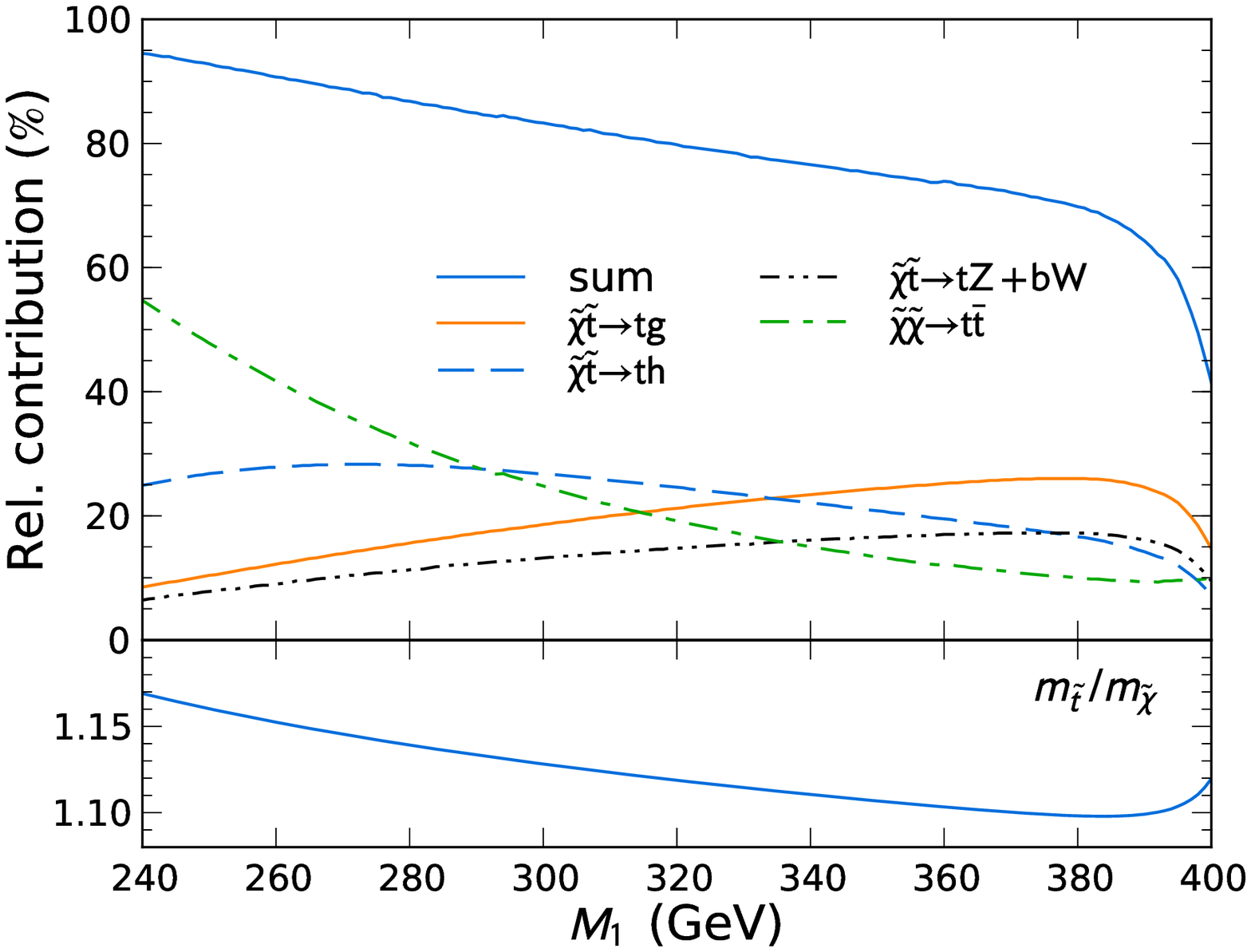}
		\includegraphics[scale=0.435]{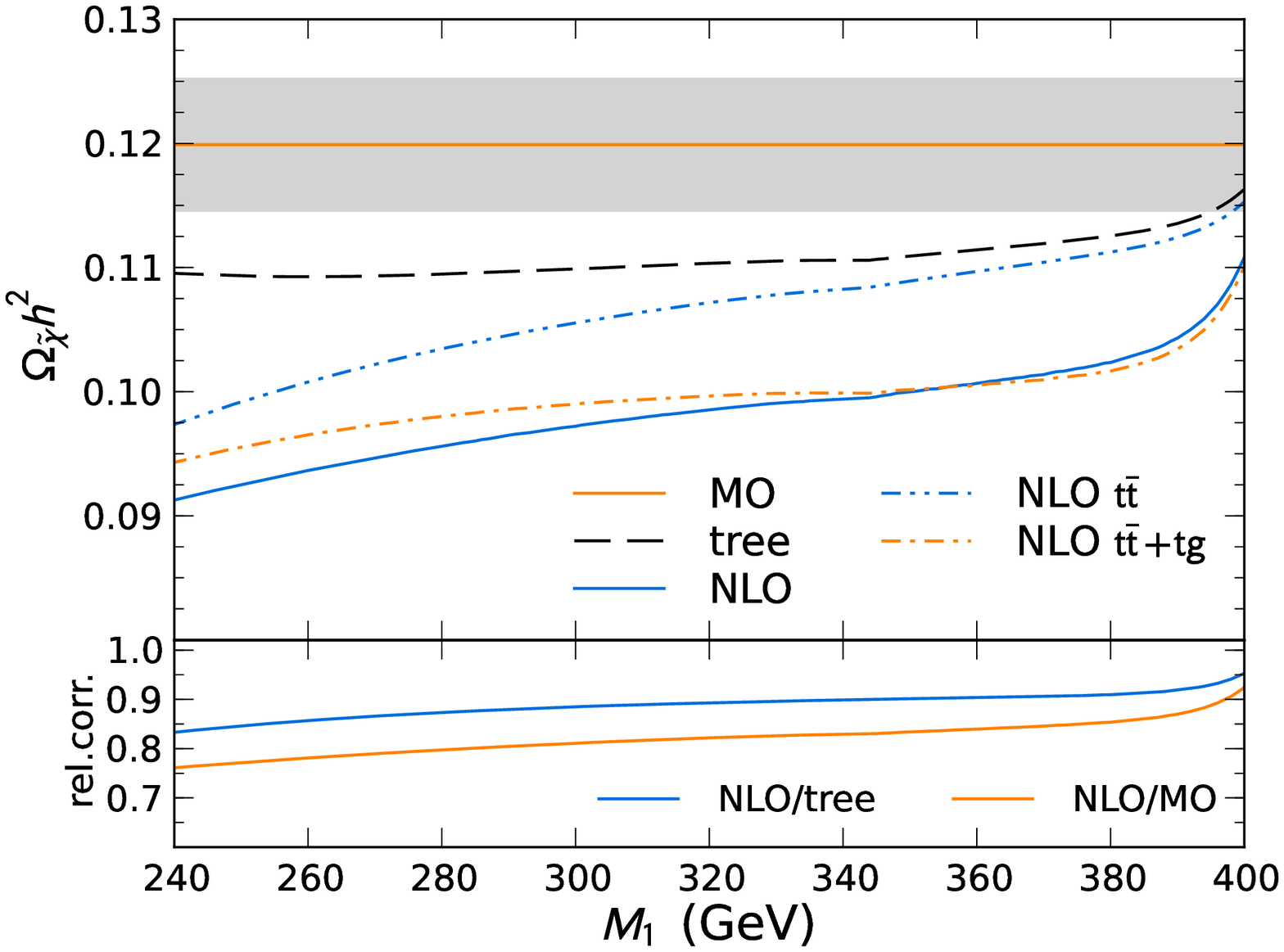}
	\end{center}
	\caption{Relative contribution of the dominant (co)annihilation channels (left) and neutralino relic density (right) along a slope in the $M_1$--$M_{q_3}$ plane in the vicinity of our example scenario. The slope is chosen such that the neutralino relic density (orange solid line) obtained by the standard {\tt micrOMEGAs} calculation exactly meets the central value of the limit given in Eq.\ (\ref{Eq:Planck}). In the right panel we show the relic density obtained by {\tt micrOMEGAs} (MO), by our tree-level calculation of the relevant processes, and by our one-loop calculation (NLO). We also indicate the relic density when taking into account one-loop corrections only for the $t\bar{t}$ final state, and only for $t\bar{t}$ and $tg$ final states. The upper and lower limits imposed by Eq.\ (\ref{Eq:Planck}) are indicated by the grey area. We show in addition the ratio between the stop and the neutralino mass (lower left) and the relative correction to the neutralino relic density (lower right).}
	\label{Fig:SlopeI}
\end{figure*}

Finally, we want to study the interplay of different channels and their corresponding corrections in our chosen scenario. 
In Fig.\ \ref{Fig:SlopeI} we study the relic density along a line in the $M_1$--$M_{\tilde{q}_3}$ plane. For each value of the bino mass parameter $M_1$, the corresponding value of the squark mass parameter $M_{\tilde{q}_3}$ is chosen such that the neutralino relic density obtained with {\tt micrOMEGAs} meets exactly the central value of the limits in Eq.\ (\ref{Eq:Planck}). 

The left plot in Fig.\ \ref{Fig:SlopeI} shows the relative contributions of all relevant channels on the top panel and 
the ratio of the stop and neutralino masses in the bottom panel. Investigating the mass ratio, 
we see that as we move towards larger values
of $M_1$, the mass of the scalar top quark gets closer to the lightest neutralino and as expected the importance 
of the coannihilation process grows. We can identify three distinct regions in the $M_1$ parameter, each with a different composition
of contributing channels. Below we will show how this composition influences the radiative corrections to the relic density. For that 
purpose, the right plot in Fig.\ \ref{Fig:SlopeI} depicts the next-to-leading corrections along the same line in the 
$M_1$--$M_{\tilde{q}_3}$ plane broken down to different contributions.

For lower $M_1$ ($M_1 \sim 250~{\rm GeV}$), the neutralino pair-annihilation into top quarks dominates and accounts for up to 55\% of the total 
annihilation cross section. The second most important contribution in this region is the neutralino-stop coannihilation into a Higgs 
boson and a quark which can reach up to 30\%. As shown in Fig.\ \ref{Fig:XSecI}, both of these processes receive substantial 
next-to-leading order corrections. The neutralino pair-annihilation into top quarks is increased by about 35\% and the coannihilation
into a Higgs boson and a quark receives a 40\% correction with respect to \MO. As a consequence, the relic density is decreased by about 18\%.

Going towards higher values of $M_1$, pair-annihilation becomes less important, and coannihilation processes with the lightest stop dominate. This is due to the fact that the stop mass is getting closer to the neutralino mass, as illustrated in the lower part 
of the left panel in Fig.\ \ref{Fig:SlopeI}. Coannihilation dominates when the relative mass difference is lower than 15\%. 
For medium values of $M_1 \sim 320\ {\rm GeV}$, three contributions, the neutralino pair-annihilation, the neutralino-stop 
coanihilation into a Higgs boson and a quark and the neutralino-stop coanihilation into a gluon and a quark, 
compete and each of them amount to about 20\%. Although all three processes contribute almost the same, the radiative corrections 
to these processes are not the same at all. Out of the processes considered, the largest corrections come from the coannihilations with
the strongly interacting gluon in the final state. Even though, the corrections to the processes are large, the total correction to the 
relic density is not as large as for small $M_1$ because we are correcting only 70\% of the total annihilation cross section as compared to 85\% in the case of lower values of $M_1$.  

For large values of $M_1$ ($M_1\sim 380\ {\rm GeV}$), the coannihilation into a quark and a vector boson takes over. Out of all vector boson final states, the gluon is enhanced the most due to the strong coupling constant. The next important channel is the coannihilation 
into $bW^+$ which is larger than the rest because the bottom quark is much lighter than the top quark. The neutralino pair-annihilation
is largely suppressed and the coannihilation with $th^0$ in the final state is reduced as well, although it is still comparable with the coannihilation into $bW^+$. These facts are also reflected in the decomposition of the next-to-leading corrections to the relic density
(shown on the right in Fig.~\ref{Fig:SlopeI}). Almost 90\% of the correction stems from the gluon final state and the neutralino 
pair-annihilation hardly contributes to the loop correction. The remaining two processes are of almost equal importance with corrections
of opposite signs, which means that their contribution to the total correction almost cancels and changes signs at about $M_1\sim 360\ {\rm GeV}$ as can be seen on the 
right in Fig.~\ref{Fig:SlopeI}. Although for large $M_1$ the fraction of processes we correct is not 
as large (only about 65\%), the one-loop SUSY-QCD correction is still substantial (about 15\%) owing to the large correction to the $tg$ final state.

In summary, the comparison between our one-loop result and the values obtained by {\tt micrOMEGAs}, as shown in Figs.\ \ref{Fig:2Drelic} and \ref{Fig:SlopeI}, demonstrates that with corrections to the theoretically predicted relic density of up to almost 25\% the one-loop corrections can be significant and therefore necessary to take into account for a precise determination of the favored parameter space.

% =============================================================================
%!TEX root=paper.tex

\section{Conclusions \label{Sec:Conclusion}}
% =============================================================================
In this paper, we have studied for the first time the combination of one-loop corrections to neutralino pair-annihilation and 
neutralino-stop coannihilation. We extended our previous work \cite{DMNLO_Stop1} by the coannihilation process with a gluon in the final 
state, which needs additional, dedicated treatment with respect to other neutralino-stop coannihilation processes. We have described in 
detail the performed $\alpha_s$ renormalization including the full 2-loop MSSM matching coefficients 
as well as the derivation of the gluon wave-function renormalization constant. As in the case of a gluon in the final state, 
not only soft, but also collinear divergences appear, the one-cutoff phase-space slicing as used in Ref.~\cite{DMNLO_Stop1} was extended 
by using phase-space slicing with two cutoffs. We have used the eikonal and the leading pole approximation in the soft and collinear 
limit, respectively, and have shown in detail how the poles can be extracted in this case. Further, we have demonstrated that this method 
renders the real emission finite without being cutoff dependent.

We have chosen a representative parameter point, where we have shown that a scenario with an admixture of neutralino 
pair-annihilation into quarks and coannihilation meets the today's limits regarding the relic density constraint, the Higgs mass and low 
energy observables. Further, this kind of scenarios with a light stop being almost degenerate in mass with the neutralino LSP is very 
attractive as they are not yet excluded by direct or monojet searches in this mass region.

We have discussed in detail the impact of our performed one-loop correction to the (co)annihilation cross sections for different final 
states. Corrections of up to $40-45~\%$ with respect to the default {\tt micrOMEGAs} value have been observed, 
especially for the lightest Higgs and gluon final state. Combining all corrected channels, which make up roughly $80~\%$ of the total 
(co)annihilation cross section in our example scenario, a relative correction to the default {\tt micrOMEGAs} relic density of almost 
$20~\%$ is achieved. This leads to a clear shift of the Planck one-sigma band in the parameter space. Thus, we could demonstrate that 
these corrections are non-negligible and therefore interesting to be taken into account.

All loop corrections are implemented in a computer package, called {\tt DM@NLO} \cite{DMNLO}, which can be easily linked to {\tt 
micrOMEGAs} in order to obtain a more precise theoretical prediction of the neutralino relic density.

% =============================================================================
\acknowledgments

The authors would like to thank Q.~Le~Boulc'h for his participation in the early stages of this work and M.~Meinecke and P.~Steppeler for useful discussions. We are grateful to A.~Pukhov for providing us with the necessary functions to implement our results into the {\MO} code. The work of J.H.\ was supported by the London Centre for Terauniverse Studies (LCTS), using funding from the European Research Council via the Advanced Investigator Grant 26735. This work was also supported by DAAD/EGIDE, Project No.\ PROCOPE 54366394. The work of M.K. and K.K.\ is supported by the Helmholtz Alliance for Astroparticle Physics. J.H.\ and B.H.\ acknowledge the NORDITA programme ``What is the dark matter?'', in the context of which important stages of the presented work have been completed.

% =============================================================================
%\newpage
\appendix
%!TEX root=paper.tex

%==========================================================================
\section{Details of the two-cutoff method \label{App:cutoff}}

%==========================================================================
\subsection{Soft integrals \label{App:cutoff_soft}}

The integrals for the purely soft divergent cases
\begin{align}
	I_{ab}= \int_{|\vec{k}|\leq\delta_s} \frac{d^{3}k}{2 \omega}\left[ \frac{2 a.b}{(a.k)(k.b)} \right]
	\label{Eq:Iabansatz}
\end{align}
can be generically calculated. Denoting $\Delta = \frac{1}{\varepsilon} - \gamma_E + \log 4\pi$, the result for the self-contracted case is
\begin{align}
	I_{a^2}= 2 \pi \left\{ -\Delta + \mathrm{log}\frac{4\delta_s^2}{\mu^2} + \frac{a^0}{|\vec{a}|} \log \frac{a^0 - |\vec{a}|}{a^0 + |\vec{a}|} \right\}.
\label{Eq:Iaafinal}
\end{align}
The corresponding integral for an interference of two real emission diagrams with different radiated particles $a$ and $b$ can be written as
\begin{widetext}
\begin{align}
	I_{ab} = \frac{4 \pi \alpha (a . b)}{(\alpha a)^2 - b^2} & \left\{ \vphantom{ \frac{P^0 - |\vec{P}|}{P^0 + |\vec{P}|}} \frac{1}{2} \! \left( \! -\Delta + \mathrm{log}\frac{4\delta_s^2}{\mu^2} \right) \log \frac{(\alpha a)^2}{b^2} \right.  \nonumber \\
	& + \left. \left[ \frac{1}{4} \log^2 \frac{P^0 - |\vec{P}|}{P^0 + |\vec{P}|} + \mathrm{Li}_2 \!\! \left( \! 1 -\frac{P^0 - |\vec{P}|}{\frac{\alpha^2 a^2 - b^2}{2 (\alpha a^0 -b^0)}} \! \right) +  \mathrm{Li}_2 \!\! \left( \! 1 -\frac{P^0 + |\vec{P}|}{\frac{\alpha^2 a^2 - b^2}{2 (\alpha a^0 -b^0)}} \! \right)\right]_{P=b}^{P=\alpha a} \right\},
	\label{Eq:Iabfinal}
\end{align}
\end{widetext}
with
\begin{align}
	\alpha = \frac{2 p_i p_j \pm \sqrt{4 (p_i.p_j)^2 - m_i^2 m_j^2} }{2m_i^2},
\end{align}
where $m_i$ and $m_j$ describe the masses of the particles of which a gluon is radiated off. The symbols $p_i^\mu$ and $p_j^\mu$ indicate their corresponding momenta. The solution is chosen such that the condition $(\alpha a^0 - b^0)/b^0 > 0$ is valid. A detailed derivation of the above integrals can be found in Ref.\ \cite{Veltmann1979}. 

%==========================================================================
\subsection{Definition of momenta and energies \label{App:cutoff_momenta}}

In the following, we summarize the definitions of the $D$-dimensional momenta of the real emission process. We follow the notation of Ref.\ \cite{HarrisOwens}.

In the soft-collinear case, the momentum of the additionally radiated gluon or the light quark is defined as
\begin{align}
	k^\mu &= k\left(1,...,0,\sin \theta_1\sin \theta_2, \sin \theta_1 \cos \theta_2, \cos \theta_1\right).
\end{align}
The $D$-dimensional momenta of the corresponding tree-level diagrams can be written as
\begin{align}
	p_1^\mu &= \frac{\sqrt{s}}{2}\left(\frac{2 E_1}{\sqrt{s}},...,0,0,\beta_1\right),\\
	p_2^\mu &= \frac{\sqrt{s}}{2}\left(\frac{2 E_2}{\sqrt{s}},...,0,0,-\beta_1\right),\\
	p_3^\mu &= \frac{\sqrt{s}}{2}\left(\frac{2 E_3}{\sqrt{s}},...,0,\beta_2 \sin \theta,\beta_2 \cos \theta \right),\\
	p_4^\mu &= \frac{\sqrt{s}}{2}\left(\frac{2 E_4}{\sqrt{s}},...,0,-\beta_2 \sin \theta,-\beta_2 \cos \theta \right),
\end{align}
where $p_1^\mu$ and $p_2^\mu$ indicate the momentum of the incoming particles (e.g.\ neutralino and stop) and $p_3^\mu$ and $p_4^\mu$
the outgoing particles (e.g.\ top and gluon). Their energies and velocities $\beta_{i=1,2}$ are given by
\begin{align}
	E_1 &= \frac{s + m_1^2 - m_2^2}{2 \sqrt{s}} \quad \quad E_2 = \frac{s + m_2^2 - m_1^2}{2 \sqrt{s}},\\
	E_3 &= \frac{s + m_3^2 - m_4^2}{2 \sqrt{s}} \quad \quad E_4 = \frac{s + m_4^2 - m_3^2}{2 \sqrt{s}},\\
	\beta_1 &= \frac{\lambda^{1/2}(s,m_1^2,m_2^2)}{s} \quad \beta_2 = \frac{\lambda^{1/2}(s,m_3^2,m_4^2)}{s},
\end{align}
with $\lambda(x,y,z)$ being the K\"all\'en function.

In the collinear limit ($\vec{p}_t^2 \ll (zp)^2$), we can express the outgoing momenta of the two relevant particles $p_4^\mu$ and $p_5^\mu$ with one effective momentum $p_{45}^\mu = p_4^\mu + p_5^\mu + \mathcal{O}(p_t^2)$
\begin{align}
	p_{45}^\mu &=\left(p,0,0,p\right)\\
	p_4^\mu &\simeq \left( zp + \frac{\vec{p}_t^2}{2zp},\vec{p}_t, zp  \right)\\
	p_5^\mu &\simeq \left( (1-z)p + \frac{\vec{p}_t^2}{2(1-z)p},-\vec{p}_t, (1-z)p  \right),
	\label{Eq:momentahardcoll}
\end{align}
where $\vec{k}_t$ indicates the transverse components of the particle with momentum $p_4^\mu$ in the centre-of-mass system. Also, $z$ denotes the momentum fraction of particle $p_4$ with respect to the merged $D$-dimensional momentum $p_{45}$ and $(1-z)$ the complementary fraction of particle $p_5$ in the direction of the z-axis.

%==========================================================================
\subsection{Soft-collinear integrals \label{App:cutoff_softcoll}}

In the soft-collinear case it is advantageous to re-write the corresponding scalar products such that they obey the form
\begin{align}
	I^{(l,m)}_\varepsilon &= \int_0^\pi d\theta_1 \sin^{1 - 2\varepsilon}\theta_1\,\,\int_0^\pi d\theta_2 \sin^{-2\varepsilon}\theta_2\nonumber\\
		&\times \frac{(a + b \cos \theta_1)^{-l}}{(A + B \cos \theta_1 + C \sin \theta_1 \cos \theta_2)^{m}}.
		\label{Eq:genericsoftcollintappendix}
\end{align}
Most of these integrals are well-known and tabulated in the literature, see e.g.\  Refs.\ \cite{Beena1989, vanNeerven1986, Smith1989, Harris1995}. According to the relation between $A$, $B$, and $C$ as well as $a$ and $b$, special integrals have to be chosen. In our case, we need two different integrals. The first one,
\begin{align}
	I^{(0,1)}_\varepsilon &= \frac{\pi}{\sqrt{B^2 + C^2}}\left\{  \ln \frac{A + \sqrt{B^2 + C^2}}{A - \sqrt{B^2 + C^2}} \right. \label{Eq:I01}\nonumber\\
&+ 2 \left. \varepsilon \left[ \mathrm{Li}_2 \left( \frac{2 \sqrt{B^2 + C^2} }{A + \sqrt{B^2 + C^2} } + \frac{1}{4} \ln^2 \frac{A + \sqrt{B^2 + C^2}}{A - \sqrt{B^2 + C^2}} \right) \right]\right.\nonumber\\
&+\left. \mathcal{O}(\varepsilon^2) \right\},
\end{align}
fulfills the condition $A^2 \neq B^2 + C^2$ and can be found in Ref.\ \cite{Harris1995}.

For the second one a more general integral is necessary, which corresponds to the conditions $a^2 \neq b^2$ as well as $A^2 = B^2 + C^2$. To render finite contributions, we need this type of integral up to $\mathcal{O}(\varepsilon)$, which could not be found in present literature. Therefore, a similar integral of Ref.\ \cite{Bojak2000} has been used for its derivation,
\begin{align}
	I^{(1,1)}_\varepsilon &= \pi \frac{1}{aA - bB} \left\{ - \frac{1}{\varepsilon} + \ln \frac{(aA - bB)^2}{(a^2 - b^2)A^2}\right. \nonumber\\
&\left. - \varepsilon \left[ \ln^2 \left( \frac{(a-b)A}{Aa - bB} \right) - \frac{1}{2}\ln^2 \left( \frac{a +b}{a - b} \right) \right. \right. \nonumber \\
&+ \left. \left.  2 \mathrm{Li}_2 \left( \frac{b (B-A)}{A (a-b)} \right) - 2 \mathrm{Li}_2 \left( \frac{-b (A + B)}{Aa - bB} \right)\right]\right. \nonumber\\
&\left. + \mathcal{O}(\varepsilon^2) \right\}.
 \label{Eq:I11}
\end{align}
We have verified the corresponding terms up to $\mathcal{O}(\varepsilon^0)$ by comparing with Ref.\ \cite{Smith1989}.

%==========================================================================
\subsection{Hard-collinear integrals \label{App:cutoff_hardcoll}}

In the collinear limit the squared matrix element of a $2 \rightarrow 3$ process factorizes due to the factorization theorem \cite{Collins1985, Bodwin1984} into the leading order squared matrix element and an Altarelli-Parisi splitting kernel \cite{AltarelliParisi}. For the expressions appearing in this paper, we need the $D$-dimensional unregulated splitting functions
\begin{align}
P_{ij}(z,\varepsilon) = P_{ij}(z) + \varepsilon P^{\prime}_{ij}(z)
\end{align}
with \cite{HarrisOwens}
\begin{align}
	P_{gg}(z) &= 2N \left[ \frac{z}{1-z} + \frac{1+z}{z} + z(1-z) \right]\,,\\
	P^\prime_{gg}(z)&=~0\,,\\
	P_{qg}(z) &= \frac{1}{2} \left[ z^2 + (1-z)^2 \right]\,,\\
	P^\prime_{qg}(z)&=-z (1-z)\,.
\end{align}

The integration bounds of Eq.\ (\ref{Eq:integrationboundsgg}) can be obtained by applying the hard condition on the energies $E_4$ and $E_5$. With the hard condition
\begin{align}
\delta_s \frac{\sqrt{s_{12}}}{2} \leq E_{4,5} \leq \frac{\sqrt{s_{12}}}{2} \left( 1 - \frac{m_3^2}{s_{12}} \right)
\end{align}
and $E_4$ and $E_5$ being defined as
\begin{align}
	E_5 = \frac{s_{12} - s_{34}}{2 \sqrt{s_{12}}} \quad \mathrm{and} \quad E_4 = \frac{s_{12} - s_{35}}{2 \sqrt{s_{12}}}, 
\label{Eq:energy45}
\end{align}
the integration bound for $z$ can be derived as 
\begin{align}
	1 - \frac{1 - \frac{\delta_s}{\beta}}{1 - \frac{s_{45}}{s_{12}}\frac{1}{\beta}} 
	\leq z \leq 1 - \frac{1 - \frac{\delta_s}{\beta}}{1 - \frac{s_{45}}{s_{12}}\frac{1}{\beta}},
\end{align}
with $\beta$ being defined as
\begin{align}
	\beta = 1 - \frac{m_3^2}{s_{12}}.
\end{align}
The relations necessary for the derivation are:
\begin{align}
	s_{12} &= (p_3 + p_{45})^2 \simeq m_3 ^2 + 2 p_3. p_{45} + s_{45}\,,\nonumber \\
	s_{34} &=(p_3 + p_4)^2 = m_3^2 + 2p_3.p_4\nonumber \\
		&\simeq m_3^2 + z(2 p_3. p_{45})\nonumber \\
		&\simeq m_3^2 + z(s_{12} - m_3^2 - s_{45})\,,\nonumber \\
	s_{35} &=(p_3 + p_5)^2 = m_3^2 + 2p_3. p_5\nonumber \\
		&\simeq m_3^2 + (1 - z)(2 p_3.p_{45})\nonumber \\
		&\simeq m_3^2 + (1 - z)(s_{12} - m_3^2 - s_{45}).
\label{Eq:s34s35approx}
\end{align}
With these conditions the $\mathcal{O}(\frac{\delta_c}{\delta_s})$ term can be retained as described in Ref.\ \cite{HarrisOwens}.

% =============================================================================
%!TEX root=paper.tex
\bibliographystyle{apsrev}

\end{document}